\documentclass{article}
\usepackage[utf8]{inputenc}
\usepackage[english]{babel}

\usepackage{amsmath,amsthm,amssymb}
\usepackage{dsfont}
\usepackage{graphicx}
\usepackage{hyperref}

\usepackage{fullpage}

\theoremstyle{definition}
\newtheorem{theorem}{Theorem}
\newtheorem{definition}{Definition}

\newtheoremstyle{mystyle}  
  {}                       
  {}                       
  {\normalfont}           
  {}                      
  {\bfseries}             
  {: }                     
  { }                     
  {}                      

\theoremstyle{mystyle}
\newtheorem*{bc}{Bell's Conclusion}

\newcommand{\refbc}[1]{\hyperref[#1]{Bell's Conclusion}}

\usepackage{enumitem}

\usepackage[authoryear, round]{natbib}

\title{What are the bearers of hidden states? On an important ambiguity in the formulation of Bell's theorem}
\author{Joanna Luc}

\newcommand*{\inferencel}[3][t]{%
   \begingroup
   \def\and{\\}%
   \begin{tabular}[#1]{@{\enspace}l@{\enspace}}
   #2 \\
   \hline
   #3
   \end{tabular}%
   \endgroup
}

\usepackage{setspace}

\usepackage{hyperref}

\usepackage{float}

\begin{document}

\maketitle

\begin{abstract}
\noindent
One of the conclusions that Bell drew from his famous inequality was that any hidden variable theory that satisfies Local Causality is incompatible with the predictions of Quantum Mechanics for Bell's Experiment. However, Local Causality does not appear in the derivation of Bell's inequality. Instead, two other assumptions are used, namely Factorizability and Settings Independence. Therefore, in order to establish the mentioned Bell's conclusion, we need to relate these two assumptions to Local Causality. The prospects for doing so turn out to depend on the assumed location of the hidden states that appear in Bell's inequality. In this paper, I consider the following two views on such states: (1) that they are states of the two-particle system at the moment of preparation, and (2) that they are states of thick slices of the past light cones of measurements. I argue that straightforward attempts to establish Bell's conclusion fail in both approaches. Then, I consider three refined attempts, which I also criticise, and I propose a new way of establishing Bell's conclusion that combines intuitions underlying several previous approaches.
\end{abstract}

\section{Introduction}

\citet{epr} famously argued that Quantum Mechanics (QM) cannot be regarded as a complete theory---there should exist some additional (``hidden'' in the sense of not being captured by the quantum state) variables that explain in a local way the correlations that occur in quantum experiments. A major contribution to the investigation of theories postulating such variables (called ``hidden variable theories'', in short ``HVTs'') was made by \citet{Bell1964}, who proposed a general framework in which one can analyse all such theories at once. Within this framework, he derived an inequality (called ``Bell's inequality'' or ``Bell's theorem'') that constrains the expectation values for the results of a certain spin measurement experiment (which is now called ``Bell's Experiment''). Bell's inequality is violated by QM predictions for Bell's Experiment, but it is satisfied by a certain class of HVTs, which Bell identified with the class of local HVTs. Therefore, Bell's own interpretation of his result was that it allows us to exclude all local HVTs by virtue of their being inconsistent with the predictions of QM for Bell's Experiment. Since the results of actual realisations of Bell's Experiment turned out to confirm QM predictions and violate Bell's inequality, this interpretation---if correct---would lead to the conclusion that Nature itself is nonlocal, contrary to what Einstein expected (cf. \citealt{norsen-bell-jarrett-2009} and \citeyear{norsen-bell-2011}, \citealt{maudlin-what-bell-did}).

Locality is understood here in a way closely related to relativity theory: roughly, as the claim that no event can be causally influenced by anything from the outside of its past light cone. However, the exact formulation of this principle is a subtle issue. In his last paper, Bell's preferred formulation of locality was the principle he called ``Local Causality'' (see section \ref{subsec:state-of-region-Bell} for details), so his conclusion from his theorem can be stated as follows:

\begin{quote}
\begin{bc}
\phantomsection
\label{bell}
All HVTs that satisfy Local Causality are inconsistent with the predictions of QM for Bell's Experiment.%
\footnote{In this paper, I will use the expression ``Bell's Conclusion'' as a label for this statement. This is \emph{not} meant to suggest that this is the only conclusion that Bell derived from his theorem; cf. the last paragraph of section \ref{sec:conclusion}. }
\end{bc}
\end{quote}

However, Local Causality is not used in the proof of Bell's inequality. Instead, two other assumptions are used: Factorizability and Settings Independence (see section \ref{sec:assumptions-Bell-theorem} and Appendix \ref{app:derivation-inequality} for details). Therefore, in order to retain \refbc{bell}, we need to relate Local Causality to these two assumptions. In this paper, I will investigate the prospects of doing this; we will see that this task is far from trivial. 

This paper is organised as follows. After introducing some basic concepts and notation (section \ref{sec:basic}), I will review two examples of HVTs (section \ref{sec:true-spin}) and provide a typology of such theories (section \ref{sec:typology}), in which of special interest will be locally deterministic theories, a distinguished subclass of locally causal theories. Next, I will recall the standard presentation of the assumptions of Bell's theorem (section \ref{sec:assumptions-Bell-theorem}). The central topic of this paper will be the question of what exactly the hidden states invoked in Bell's theorem encompass. The two main options considered in the literature are the hidden state of the two-particle system at the moment of preparation (section \ref{sec:state-of-system}) and the hidden state of thick slices of the past light cones of measurements (section \ref{sec:state-of-region}). In both cases, there are certain obstacles to establishing \refbc{bell}. However, there are some upgraded versions of both approaches that aim to overcome these obstacles, which will be the subject of section \ref{sec:improve}. My own proposal, presented in section \ref{sec:new-proposal}, will build upon the insights of both approaches. Section~\ref{sec:conclusion} will provide a summary of the results of this paper.

One terminological remark is in place here.
The terms ``hidden variables'' and ``hidden states'' can be misleading in at least three ways. 
First, the word ``hidden'' suggests that the postulation of these variables/states does not lead to any observable predictions. Although this might be the case for some HVTs, this is by no means a defining feature of hidden variables/states.
Second, one might suppose that HVTs do not exhaust all theories that are alternatives to the standard QM. The argument for this claim would be that some such theories may not postulate hidden variables/states. However, the word ``hidden'' here merely indicates that the postulated variables/states involve something that is not captured (or at least not fully captured) by the quantum state and either supplements or replaces it. Clearly, every theory that attempts to account for the same phenomena as the standard QM but differs from it at the level of ontology must postulate some hidden variables/states \textit{in this sense}. 
Third, the formulation of Bell's theorem in terms of HVTs may misleadingly suggest that we can avoid the conclusion of nonlocality by rejecting the assumption of the existence of hidden variables/states. However, this reasoning ignores the fact that the standard QM itself is nonlocal in some sense (although this issue is subtle; see section \ref{sec:locally-determinate}), so if there is any way to account for quantum phenomena in a local way, it requires some hidden variables/states in our sense.
Despite these confusing connotations of the word ``hidden'', I use the terms ``hidden variables'' and ``hidden states'' because this terminology has become widespread, and it makes the connection of this paper with the literature on Bell's theorem more conspicuous.%
\footnote{One of the reviewers of this journal is still dissatisfied with my usage of the terminology of ``hidden variables'' and ``hidden states'', even after these clarifications. This reviewer suggests that all instances of the word ``hidden'' are superfluous and that ``what's being described and discussed is always just a totally generic theory with whatever properties are explicitly mentioned''. I do not quite agree: some theories may describe phenomena that do not occur in Bell's Experiment (and other similar experiments), and such theories could be both local and consistent with the predictions of QM for Bell's Experiment (by virtue of not asserting anything about the latter). Therefore, an HVT is not just any theory but, as I put it in the main text, ``a theory that attempts to account for the same phenomena as the standard QM but differs from it at the level of ontology'' by postulating ``something that is not captured (or at least not fully captured) by the quantum state''. }

\section{Some basic concepts and notation}\label{sec:basic}

Since the concern with locality is motivated by relativity theory, our stage will be a spacetime with the Minkowski metric, which is used to define the light cone structure in the standard way. Additionally, we will refer to some fixed slicing of this spacetime (given by simultaneity hypersurfaces in some chosen inertial reference frame) and associated time parameter $t$. If an HVT is relativistic, in the end all its physical results must be independent of the choice of slicing.

Let $R$ be a spatiotemporal region (i.e., an open subset of $\mathbb{R}^4$) that is bounded, and let ``$t < R$'' mean that time $t$ is below $R$ (i.e., the spatial slice at $t$ lies below any spatial slice that intersects $R$). We define $\Sigma_{R, t}$ as the intersection of the spatial slice at $t$ with the past (if $t<R$) or future (if $R < t$) light cone of $R$. If $t < R$, then for any $t'<t$ we also define $C(\Sigma_{R, t}, \Sigma_{R, t'})$ to be the part of the past light cone of $R$ that lies between the surfaces $\Sigma_{R, t}$ and $\Sigma_{R, t'}$ (such regions will be informally called ``thick slices'').

We will be interested in accounting for Bell's Experiment, in which a pair of particles is prepared in an entangled spin state
\begin{equation}\label{eq:entangled-state}
\psi_{12} = \frac{1}{\sqrt{2}} \left(\psi_1^+  \psi_2^- - \psi_1^-  \psi_2^+ \right).
\end{equation}
Then, each of these particles moves from the source to one of two remote detectors, where spin in a chosen direction is measured on each of these particles. We will assume that one of two choices of direction can be made at each station (but the allowed choices do not need to be the same for both stations). The variable $a$ will range over such possible choices in the left station, whereas $b$ will range over the possible choices in the right station. Analogously, the variable $A$ will range over possible outcomes of the measurement of spin in direction $a$, and $B$ will range over possible outcomes of the measurement of spin in direction $b$ (the values of the outcomes are, by convention, $\pm 1$). We will denote the different choices of measurement settings with primes: $a' \neq a$ and $b' \neq b$; relatedly, $A'$/$B'$ will range over the possible outcomes of spin measurement in directions $a'$/$b'$. By $R_a$ and $R_b$, we will denote the spatiotemporal regions in which the respective choices of settings are made, whereas by $R_A$ and $R_B$ we will denote the spatiotemporal regions in which the measurements are performed and the outcomes are obtained (in a single run of the experiment). An additional assumption is that $R_a$ and $R_A$ are spacelike related to $R_b$ and $R_B$ (and, of course, $R_a$/$R_b$ is in the causal past of $R_A$/$R_B$).

We will be interested in the following question posed by Bell: what constraints on possible HVTs can we derive from QM predictions for Bell's Experiment? For this purpose, we will introduce hidden states $\lambda$ (the space of such states will be denoted by $\Lambda$), which might be different for different HVTs. Such $\lambda$ encompasses all hidden variables considered in a given theory.
For $\lambda$, Bell also used the term ``the beables of the theory'', which he defines as ``those entities in [the theory] which are, at least tentatively, to be taken seriously, as corresponding to something real'' \citep[p.~234]{Bell1990}.%
\footnote{All page references to Bell's papers are from the 2004 reprint.}
In other words, beables are ``what some candidate theory posits as being physically real'' \citep[p.~279]{norsen-bell-jarrett-2009}, without the assumption that this theory is correct.

\begin{figure}
\centering
    \includegraphics[width=0.7\textwidth]{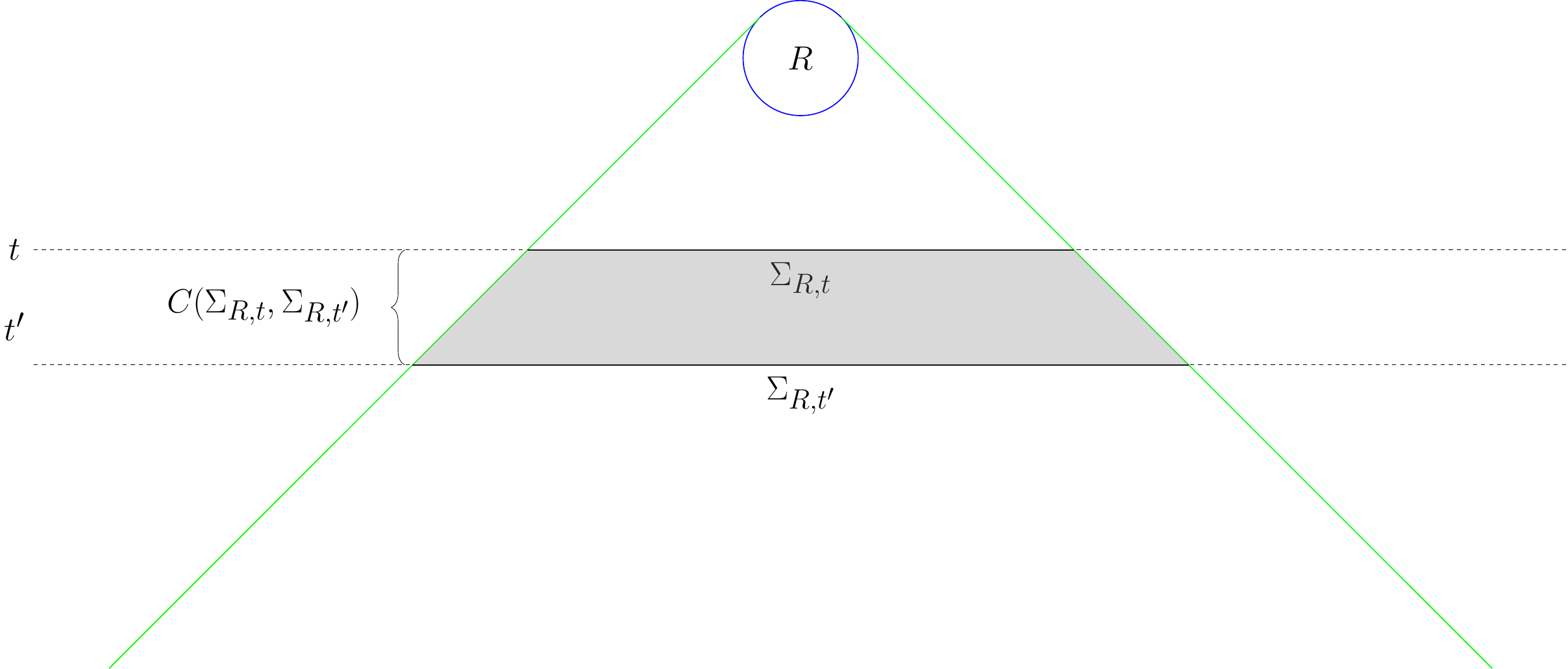}\\
    \includegraphics[width=0.7\textwidth]{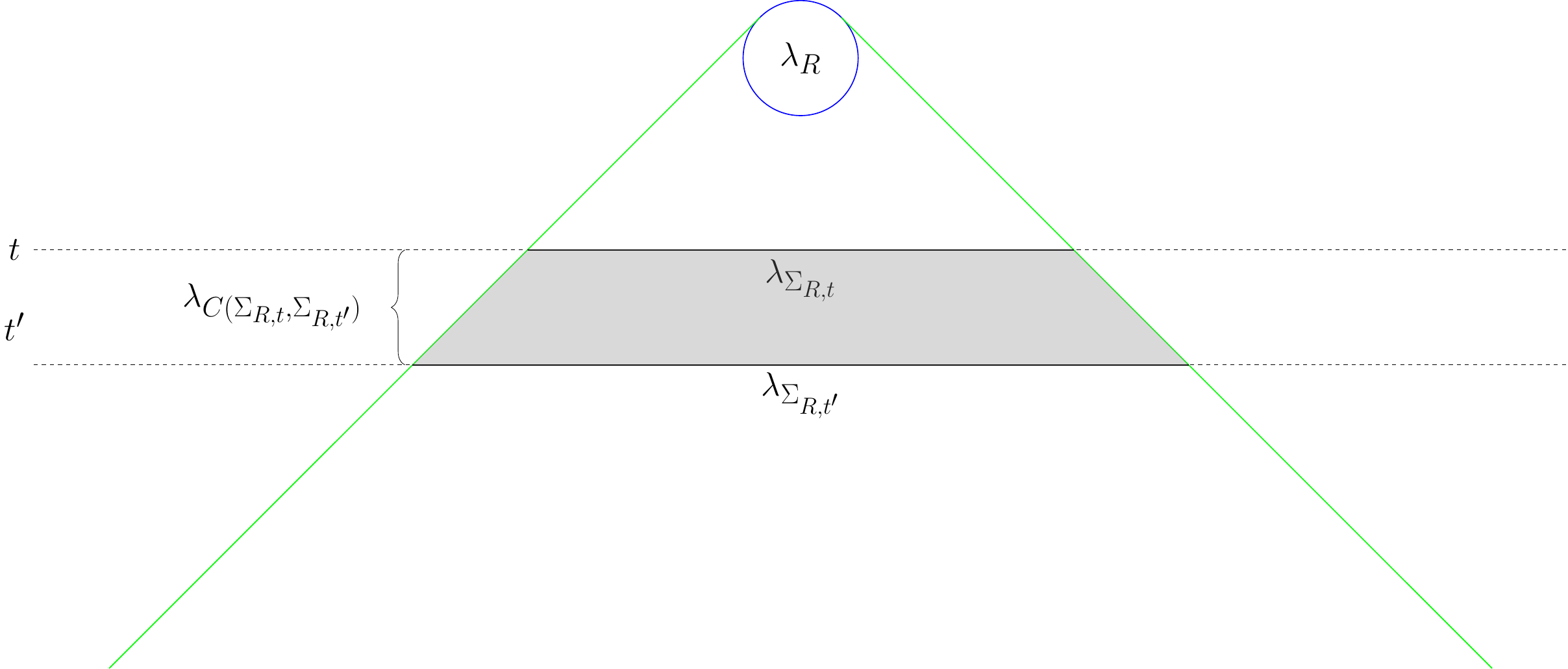}
\caption{My notation for regions (top) and states (bottom); $C_{\Sigma_{R, t}, \Sigma_{R,t'}}$ covers only the shaded region.}
\label{fig:notation}
\end{figure}

We will later need to ascribe such states to various regions of spacetime and space. For any spatiotemporal region $R$, by $\lambda_R$ we denote a possible complete specification of hidden variables (of a given HVT) in $R$. Different such possible specifications will be denoted with primes (i.e., our convention is that $\lambda'_R \neq \lambda_R$). Analogously, we can talk about complete specifications of hidden variables on spatial slices (denoted by $\lambda_{\Sigma_{R, t}}$) or on thick slices (denoted by $\lambda_{C(\Sigma_{R, t}, \Sigma_{R, t'})}$). Finally, $\lambda_t$ means the hidden state of the entire space at time $t$. My notation for regions and states is illustrated in Fig. \ref{fig:notation}.

\section{Two examples of HVTs: True Spin Theory and Bohmian mechanics}\label{sec:true-spin}

The simplest and most natural candidate for hidden variables in the case of spin measurements is just the ``true'' values of spin in every direction. As far as I know, the theory with such hidden variables has not received a name in the literature; I will call it ``True Spin Theory''. According to True Spin Theory, when we describe the spin in direction $a$ of a pair of particles by means of an entangled quantum state (such as \eqref{eq:entangled-state}), which does not ascribe any definite value of spin in direction $a$ to any of these particles, then this description is incomplete. As such, it should be completed by ascribing to each of these particles a hidden state $\lambda$ that consists of this particle's values of spin in all directions. In particular, if we consider measurements of spin in directions $a, a', b $ and $b'$, then the part of $\lambda$ that is relevant to determining the outcomes of these measurements will consist of four numbers corresponding to the values of spin in these four directions. 

True Spin Theory can be shown to be inconsistent with the predictions of QM (see, e.g., \citealt[pp.~215--227]{norsen-foundations-2017}); the reasoning used in this proof is generalised in the proof of Bell's theorem. It should be stressed that True Spin Theory, as presented here, is not really a fully developed theory. For example, we have not said anything about how definite values of spin possessed by particles might change in interactions (if at all)---that is, we have not introduced any dynamical laws. For this reason, True Spin Theory should be regarded not as a single theory but as a family of possible theories that account for Bell's Experiment in the way described above. The mentioned reasoning shows that this entire class of HVTs can be excluded as inconsistent with the predictions of QM.

However, True Spin Theory is not the only possible HVT. Another example of an HVT is Bohmian mechanics (see, e.g., \citealt{sep-qm-bohmian}), which, in contrast to True Spin Theory, is a fully developed theory with specified ontology and dynamical laws. In Bohmian mechanics, hidden variables are the true positions of particles. In the standard QM, the quantum state only provides probabilities for finding a particle in a given region of space; so, if QM is a complete theory, then we need to abandon the idea that all particles at all times have definite positions, as was presupposed by classical physics. However, this idea of classical physics is restored by Bohmian mechanics, according to which the quantum state guides how particles move, but any particle at any time has a definite position and not only a probability of being found in that position. The dynamics of this theory consists of the Schrödinger equation of the standard QM and an additional guiding equation which prescribes how the quantum state influences the positions of particles.

These are only two examples of HVTs, but we can use them to illustrate the crucial point of Bell's enterprise. Bohmian mechanics is consistent with the predictions of QM, but it is known to be nonlocal: if two particles are entangled, then what happens with one of them can have an immediate effect on what happens with the second one. (Of course, this is not a precise formulation of what nonlocality means, on which more later.) On the other hand, True Spin Theory is supposed%
\footnote{I say this in a cautious way because this is not a fully-developed theory.}
to be local: the true spins are introduced in order to account for the outcome of the measurement on one of the entangled particles without referring to the other one. However, True Spin Theory is inconsistent with the predictions of QM. The interesting question is whether there exists an HVT that has both virtues---that is, it is both local and consistent with the predictions of QM. Bell's approach promises to give an answer (a negative one) to precisely this question. Before going into the details, let us say more about different types of HVTs.

\section{A typology of HVTs}\label{sec:typology}

In this section, I will distinguish some important types of HVTs: locally determinate vs. locally indeterminate, deterministic vs. indeterministic, and locally deterministic.

\subsection{Locally determinate and locally indeterminate HVTs}\label{sec:locally-determinate}

Our notation for hidden states presupposes that they are locally determinate: if we can write $\lambda_R$ for any spatiotemporal region $R$, this means that for any such region a hidden state in that region is well-defined. However, not all theories are like this. In particular, QM states are not locally determinate: if a pair of particles is in a state \eqref{eq:entangled-state}, then we cannot ascribe any spin state to one of these particles considered in isolation. Therefore, if an HVT includes $\psi$ in an ineliminable way, then its ontology as a whole is locally indeterminate, even if the $\lambda$-part of that ontology is locally determinate.

This means that we can distinguish two types of HVTs (and, more generally, two types of theories). (1) Locally determinate theories are those theories that satisfy the following condition: for any spatiotemporal region $R$ and any possible state of that region, the restriction of that state to any subregion of $R$ is well-defined. 
In locally determinate HVTs, $\psi$ must be eliminable or at least excluded from the variables of the theory, and all the physical content of the theory must be captured by (locally well-defined) $\lambda$'s, which might then be called ``complete states''. (2) Locally indeterminate theories are those theories that are not locally determinate. 
Any HVT that retains $\psi$ as a part of its ontology is locally indeterminate.

For example, Bohmian mechanics is locally determinate if considered as a theory in which the only real physical quantity is the positions of particles (see, e.g., \citealt{esfeld-ontology-bohmian}). However, Bohmian Mechanics might also be conceived of as a theory in which particles are described by both their true position and the QM state, in which case it would be an example of a locally indeterminate HVT.

Where should we locate the standard QM in this picture? 
It seems that two variants of the standard QM need to be distinguished that differ in the ontological status they assign to the quantum state. If the quantum state is regarded as a beable of the standard QM, then this theory is locally indeterminate. This route is taken, among others, by \citet[p.~127]{norsen-foundations-2017}, who emphasises that being locally indeterminate is a stronger kind of nonlocality than causal nonlocality: 
\begin{quote}
One might think that, by showing that we
cannot cleanly apply Bell's formulation of locality to diagnose [the standard QM] as nonlocal in the EPR scenario, we leave the door open to the claim that perhaps the theory
is, after all, consistent with relativistic local causality. But that is wrong, for several
reasons.
[...] it would probably be most accurate to summarize the situation by describing quantum mechanics here as
``not even non-local''. Remember what ``locality'' means: the causal influences that
objects, moving and interacting in three-dimensional space, exert on one another,
always propagate at the speed of light or slower. A theory which fails to provide a
clear ontology of objects moving and interacting in three-dimensional space [...] doesn't even rise to the level of making the question, of whether
causal influences always propagate at the speed of light or slower, or not, meaningful.
\end{quote}

An alternative route, which excludes the quantum state from the beables of the standard QM, is taken by Bell himself. For him, the beables in the standard QM are only ``experimental procedures and experimental results'', whereas the wave function is ``a convenient but inessential
mathematical device for formulating correlations'' between such beables \citep[p.~53]{Bell1976}. 
Since the mentioned beables are locally well-defined, the standard QM, understood in this way, is locally determinate but at the price of excluding the quantum state from the ontology while retaining it in the formalism.

Summing up, we have two very different readings of the standard QM: either
\begin{enumerate}
\item[(1)] the quantum state \emph{is a beable}, in which case the standard QM \emph{is locally indeterminate} and, as suggested by Norsen, it is ``not even [causally] non-local'' or
\item[(2)] the quantum state \emph{is not a beable} and all beables are locally well-defined, in which case the standard QM \emph{is locally determinate} but not locally causal (in the sense of late Bell, see Definition \ref{def:LC-cg}), as argued by Bell (\citeyear[p.~55]{Bell1976} and \citeyear[pp.~240--241]{Bell1990}).
\end{enumerate}
Since ``the standard QM'' is a rather vague term, both of these readings are acceptable, and we should consider two variants of this theory separately.

In what follows, I will restrict my considerations to locally determinate HVTs only. 
This is justified because, as we will see, Local Causality is formulated in the framework that presupposes local determinateness (see section \ref{subsec:state-of-region-Bell} and footnote \ref{fn:nonlocal-two-senses}). 
One might then worry that, since very few HVTs considered in the literature are locally determinate, the scope of Bell's theorem is rather limited, if we assume---as I do---that this theorem does not apply to locally indeterminate theories. There are two replies to this worry. First, as we have seen, even the standard QM can be understood as a locally determinate theory if we reinterpret the role of the quantum state in a way consistent with its not being a beable. Similarly, any variant of QM that does not remove the quantum state from the formalism can be understood as either locally indeterminate (if the quantum state is included in the beables) or locally determinate (if the quantum state is excluded from the beables), provided that all other beables are locally determinate. For example, some Bohmians have proposed understanding the wave function as ``a component of physical law rather than of the reality described by the law'' \citep[p.~33]{bohmian-wave-function}, which makes this variant of Bohmian mechanics locally determinate. Second, and more importantly, the target of Bell's theorem is not the known HVTs, which are usually deliberately developed so as to be consistent with many (or even all) predictions of QM, but their potential, more local competitors. Relatedly, although the theorem may exclude some of the known HVTs, its main message is the exclusion of the entire class of HVTs that are both locally determinate and locally causal; most theories in this class will never be explicitly formulated.

\subsection{Deterministic, indeterministic and locally deterministic HVTs}

Another important division of HVTs (and, more generally, of all theories) is between deterministic and indeterministic ones. For the purpose of defining them, we will use the concept of a solution of a theory $T$, which is any specification of the variables of $T$ in the entire spacetime that is compatible with the laws of $T$, where ``compatible'' means simply ``logically consistent''. Then, we can define:
\begin{definition}\label{def:det-theory}
Theory $T$ is deterministic iff for any $t_0$ and any $\lambda_{t_0}$, there is exactly one solution that is compatible with $ \lambda_{t_0}$. Otherwise $T$ is indeterministic.
\end{definition}
\noindent 
In our terminology, ``exactly one solution that is compatible with $ \lambda_{t_0}$'' means ``exactly one specification of the variables of $T$ in the entire spacetime that is logically consistent with the laws of $T$ and $ \lambda_{t_0}$''.

The above definition is (modulo differences in formulation and pace certain subtleties, which will not be discussed here\footnote{One such subtlety is the physical equivalence of different solutions that are related by the symmetries of the theory. If a theory involves local symmetries (i.e., symmetries that are not the same at every spacetime point), such as diffeomorphisms in General Relativity and gauge symmetries in Electrodynamics, then there will be spurious witnesses of indeterminism---that is, solutions of the theory that are symmetry-related and identical on some initial segment but not identical overall. In such cases, the definition of determinism should be modified by taking into account the physical equivalence of such solutions (see, e.g., \citealt[pp.~7--9]{butterfield-hole-truth}).}) 
standard in the contemporary literature about determinism and indeterminism (see, e.g., \citealt{earman-primer} and \citealt{butterfield-det-indet}). However, it should be noted that in the literature about Bell's theorem, deterministic HVTs are often defined in a different way: namely, as theories that predict the measurement outcomes with probability $1$, given the state of the system and the measurement settings.%
\footnote{This terminology was used by Bell in later commentaries on his \citeyear{Bell1964} paper (e.g., \citeyear{Bell1971} and \citeyear{Bell1981}). However, in the 1964 paper itself, he talks only about measurement outcomes being predetermined or determined by settings together with $\lambda$; the words ``determinism'' or ``deterministic'' do not appear there.
Some other authors call determinism in this sense ``quasi-determinism'' \citep{malament-bell} or ``outcome determinism'' \citep{sep-bell-theorem}, which indicates the difference between this concept of determinism and the usual one. \label{fn:outcome-det}}
These two senses of determinism should not be conflated because, as we will see later (in the third paragraph of section \ref{sec:state-of-system}), they are not equivalent. 

An especially interesting subclass of deterministic theories is locally deterministic theories (cf. \citealt[p.~53]{Bell1976}). From the definition of determinism, it follows that for any region $R$, the physical contents of this region are uniquely determined by the full specification of the state of the world at $t < R$. However, this does not exclude the possibility that what happens in $R$ might depend on what happens at $t$ arbitrarily far away spatially from $R$, which is in tension with the relativistic idea that what happens in $R$ should not depend on what happens outside its past light cone. This leads to the concept of locally deterministic theories, in which to determine the physical contents of $R$, the information about the entire spatial slice at some $t < R$ is not needed and the part of that information that is confined to the intersection of the past light cone of $R$ with the spatial slice at $t$ suffices. More formally,

\begin{definition}\label{def:locally-det-theory}
Theory $T$ is locally deterministic iff it is locally determinate and
for any bounded spatiotemporal region $R$, for any $t_0$ such that $t_0 < R$ or $t_0 > R$
and for any $\lambda_{\Sigma_{R, t_0}}$, there exists $\lambda_{R}$ such that all solutions compatible with $\lambda_{\Sigma_{R, t_0}}$ are also compatible with $\lambda_{R}$.
\end{definition}
Since we have universal quantification over solutions, it follows from this definition that for any other $\lambda'_{R} \neq \lambda_{R}$, all solutions compatible with $\lambda_{\Sigma_{R, t_0}}$ are not compatible with $\lambda'_{R}$ (a solution on the entire spacetime cannot be compatible with different states of some region of this spacetime).
One can show that local determinism defined in this way entails determinism, so locally deterministic theories indeed form a subclass of deterministic theories (for the proof see Appendix \ref{app:loc-det-implies-det}):
\begin{theorem}\label{thm:local-det-implies-det}
Any locally determinate theory that is locally deterministic is also deterministic.
\end{theorem}

This theorem is not as trivial as its verbal formulation might suggest because ``local determinism'' has not been defined as a conjunction of ``determinism'' and ``locality'' but by a condition that does not explicitly mention determinism. Later, we will see that any locally deterministic theory is also local in the sense of satisfying Local Causality$_{C(\Sigma, \Sigma)}$ (see Theorem \ref{thm:local-det-LocalCausality} in section \ref{subsec:state-of-region-loc-det}).

There exist both deterministic and indeterministic alternatives to the standard QM. For example, GRW (see, e.g., \citealt{sep-qm-collapse}) is an indeterministic theory. True Spin Theories might be or might not be deterministic, depending on the details of their dynamics. What we know is only that they predict unique results of spin measurement given the state of the system and the settings (so they are ``outcome deterministic'', cf. footnote \ref{fn:outcome-det}). Bohmian mechanics is deterministic but not locally deterministic. This fact made this theory unattractive to Einstein, even though it was a proposal for how to complement QM, which is what he sought. This fact also led Bell to ask whether any possible completion of QM must be nonlocal. However, in order to pose this question properly, he needed to formulate a general definition of locality that can be applied to both deterministic and indeterministic theories, which he did using probability calculus (see section \ref{subsec:state-of-region-Bell}).

In order to apply any condition expressed in terms of probabilities to a deterministic theory, we would need a bridge principle that relates the two. Intuitively, it may seem that deterministic theories are extreme cases of probabilistic theories, in the sense that in the former all probabilities are $0$ or $1$. However, this relation is not so straightforward because a deterministic theory might not mention probabilities at all. For example, in classical mechanics of pointlike particles there are no probabilities; instead, there are dynamical laws in the form of differential equations, which for any initial conditions within a certain class have a unique solution. Moreover, it does not make any sense to ask what the probability of such-and-such behaviour of a particle is \textit{simpliciter}. A particle is determined to either behave in a given way or not, \textit{depending on the initial conditions}; without the full initial conditions, there is not enough information to determine how it will behave. The moral is as follows: that a theory is deterministic does not mean that any probability we can think of has value $0$ or $1$ (cf. \citealt[p.~17]{norsen-foundations-2017}). Only conditional probabilities that are conditioned on sufficient information about initial conditions can be said to have values $0$ or $1$ (although this will usually be a proper extension of the language of a deterministic theory). This leads to the following Bridge Principle:%
\footnote{What would a probability space look like in such cases? Presumably, we should regard entire solutions of $T$ as elementary events; an event corresponding to $\lambda_R$ would then be the set of all solutions that ascribe state $\lambda_R$ to region $R$. In what follows, we will write states (complete or coarse-grained) as arguments of a probability function, with the risk that such expressions will not always be well-defined.
Another technical problem is that $\lambda$ is continuous in some HVTs (e.g., in Bohmian mechanics, since particle positions are given by real numbers). This means that in many places we should use probability density instead of probability. For example, fine-grained formulation of Local Causality$_{C(\Sigma, \Sigma)}$ (i.e., Eq. \eqref{eq:local-causality}) should be in terms of probability densities, whereas coarse-grained formulation (i.e., Eq. \eqref{eq:local-causality-coarse-grained}) should be in terms of probabilities obtained by integrating the respective probability densities. For reasons of space, I will not fully do justice to this problem here.
}
\begin{definition}\label{def:bridge}
Bridge Principle Between Determinism and 0-1 Probability: 
For any theory $T$ and for any regions $R_i$ and $R_f$, if all solutions that are compatible with $\lambda_{R_i}$ are also compatible with $\lambda_{R_f}$, then $P(\lambda_{R_f} | \lambda_{R_i}) = 1$ according to $T$; analogously, if all solutions that are compatible with $\lambda_{R_i}$ are not compatible with $\lambda_{R_f}$, then $P(\lambda_{R_f} | \lambda_{R_i}) = 0$ according to $T$.
\end{definition}
Conditional probabilities of other events might be introduced in a different way (e.g., by specifying the probability density over the initial states, as in Bohmian mechanics), but they might be not defined at all.

\section{Assumptions of Bell's theorem}\label{sec:assumptions-Bell-theorem}

There are several versions of Bell's theorem, each of which involves a different inequality or set of inequalities. The most famous of them is the Bell-CHSH inequality, named after Bell, Clauser, Horne, Shimony and Holt \citep{chsh}. The proofs of these various inequalities use two assumptions, which are called Factorizability and Settings Independence (see Appendix \ref{app:derivation-inequality} for the proof of Bell-CHSH inequality):%
\footnote{Since we are considering only situations in which the experiment takes place, $P(\lambda)$ should be non-zero only for those $\lambda$ that are consistent with Bell's Experiment being performed, which entails that every $\lambda$ with non-zero probability (density) should be consistent with at least one choice of measurement settings. For this reason, we can think of $P(\lambda)$ as being a conditional probability $P(\lambda | \text{Bell's Experiment})$.}
\begin{quote}
\textbf{Factorizability}: $P(A, B | a, b, \lambda) = P (A | a, \lambda)  P(B | b, \lambda)$.
\end{quote}

\begin{quote}
\textbf{Settings Independence}: $P(\lambda | a,b) = P(\lambda) $.
\end{quote}

Starting with the latter, Settings Independence expresses the independence of the hidden state $\lambda$ from the measurement settings $a$ and $b$.%
\footnote{\label{fn:SettInd-direction} Technically, Settings Independence is a statement about the probability (density) of $\lambda$ conditioned on settings, not the other way around. 
However, this does not mean that the violation of this condition would be a causal influence of settings on the hidden state (such an influence, if it occurred, would be retrocausal because the hidden state is provided at a time \textit{before} the choice of settings). 
This is just a probabilistic independence condition, so the direction of conditionalization in the formula does not need to be aligned with the direction of causation in the case of the violation of this condition (cf. footnote \ref{fn:SI-violations}).}  
It is also called ``statistical independence'', ``measurement independence'', ``no conspiracy'' and even ``free will'' or ``free choice''. The latter two names are motivated as follows: measurement settings are (or at least can be) chosen by humans, so if these choices are free, then they should not depend on the hidden state that is measured. The hidden state being such-and-such cannot prevent us from choosing these particular settings rather than others; moreover, it should not even influence the probability of our choice of these particular settings.

Factorizability is usually regarded as derived from some more basic principles. \citet{jarrett-locality} suggested that Factorizability should be viewed as derived from two principles, which \citet{shimony-events} called ``Parameter Independence'' and ``Outcome Independence'' (Jarrett called them ``locality'' and completeness'', but Shimony's terminology has become more popular). 
Since these terms do not appear in Bell's (\citeyear{Bell2004}) collected papers, one may suppose that this view on Factorizability as deriving from Parameter Independence and Outcome Independence was never endorsed by Bell. Instead, in his last paper \citep{Bell1990}, he regards Factorizability as derived from another principle, which he calls Local Causality. Its initial formulation is as follows \citep[p.~239]{Bell1990}:
\begin{quote}
The direct causes (and effects) of events are near by, and even
the indirect causes (and effects) are no further away than
permitted by the velocity of light.
\end{quote}
Subsequently in that paper, Bell provides a more precise formulation of this principle, which will be analysed in detail in section \ref{subsec:state-of-region-Bell}.

The Bell-CHSH inequality, which follows from Factorizability and Settings Independence, turns out to be violated by the predictions of QM. Moreover, experimental tests have confirmed its violation in a way predicted by QM (starting with \citealt{aspect-bell-exp}; for an overview see \citealt{sep-bell-theorem}, sections 4 and 5). Therefore, we need to conclude that at least one of these two assumptions is false (unless we find some other assumption or presupposition of the theorem that is more suspect). Most commentators, including Bell and \citet{chsh}, regard Settings Independence as well justified on independent grounds, so their conclusion is that Factorizability is false. However, according to \citet{Bell1990}, the proper conclusion should be that Local Causality is false. The aim of the rest of this paper is to investigate how one can retain \refbc{bell}.

In the above considerations, we did not specify to what exactly the hidden state $\lambda$ is ascribed (i.e., what the bearer of this state is). It turns out that, in the literature, there are two main ways of thinking about $\lambda$: it is regarded either as a state of the two-particle system at the moment of preparation, or as a state of the thick slices of the past light cones of measurements. The next two sections, \ref{sec:state-of-system} and \ref{sec:state-of-region}, will analyse these two options in more detail.

\section{Hidden state as a state of the two-particle system at the moment of preparation}\label{sec:state-of-system}

The first option for the choice of $\lambda$ is to take it to be the state of the measured system \textit{only}. For example, \citet{sep-bell-theorem} write:
\begin{quote}
This conceptual framework first of all postulates an ensemble of pairs of systems, the individual systems in each pair being labeled as 1 and 2. Each pair of systems is characterized by a ``complete state'' $\lambda$ which contains the entirety of the properties of the pair at the moment of generation. 
\end{quote}
We will denote this state by $\lambda_{t_P, 1, 2}$, to be read as ``the hidden state of the system consisting of particles $1$ and $2$ at the time of preparation of this system for Bell's Experiment, $t_P$''.
This expression should not be understood as presupposing the existence of classical-like particles; ``the system consisting of particles $1$ and $2$'' might be a physical system of any kind that corresponds, in a given HVT, to what we ordinarily recognise as two particles on which the measurement is performed. What matters for our purposes is that $\lambda_{t_P, 1, 2}$ is the state of the measured system at the moment of preparation, which is located in the intersection of the light cones of measurements (see Fig. \ref{fig:Bell-exp}).
I prefer the term ``preparation'' over ``generation'' because the latter suggests that the particles did not exist before $t_P$, which might or might not be the case. This way of thinking about the hidden state agrees with ``early'' Bell (\citeyear[p.~15]{Bell1964} and \citeyear[p.~36]{Bell1971})
and with many of his followers and commentators, including \citet[p.~881]{chsh}, \citet[p.~130]{shimony-naturalistic}, \citet[p.~32]{van-fraassen-charybdis} and many others. It seems that identifying $\lambda$ in Bell's theorem with $\lambda_{t_P, 1, 2}$ is predominant in the literature, but this is difficult to assess due to the large size of that literature. The idea behind this choice of $\lambda$ is that the correlations between the outcomes of measurements on remote particles can be explained by the particles' interaction in the past, the results of which are encoded in the subsequent states of each of them.

The problem with this choice of $\lambda$ is that the explanation of correlations between measurement outcomes by the properties of the measured system at the time of preparation does not exhaust all conceivable local ways of explaining such correlations. Only in some special cases can we be sure that there are no other possible local explanations. In True Spin Theory, according to which hidden variables are the true values of spin in every direction, it is arguably sufficient to consider $\lambda_{t_P,1,2}$. This is the case at least if we assume that (i) the measurement of spin reveals its true value and (ii) this value does not change between the time of the preparation of the system and the time of measurement. Assumption (i) seems to be justified just by the definition of spin measurement (at least under True Spin Theory's assumption that every particle has a definite value of spin in every direction at every time), whereas (ii) follows from assuming that the system is appropriately isolated between its preparation and measurement. The latter does not mean that the state of the system does not change at all during this interval; in particular, particles are supposed to move (from the source to the detector), so their positions \textit{do} change during that interval (or rather, what changes is the quantum state that prescribes the probabilities of particles being found in various positions if measured, unless we consider a variant of True Spin Theory that postulates that particles have definite positions in addition to definite values of spin in all directions). What is required to not change between the time of preparation and the time of measurement is only the particles' true values of spin. If we accept these assumptions, then probabilities of the form $P(A | a, \lambda_{t_P,1,2})$ and $P(B | b, \lambda_{t_P,1,2})$ all have values $0$ or $1$. This is because the outcome of spin measurement should only depend on the direction in which we measure spin and the particle's true value of spin in that direction (by virtue of (i)); however, this true value is the same at the time of measurement as it was at the time of preparation (by virtue of (ii)).

However, assumptions (i) and (ii) cease to be valid if hidden variables are different from the true values of spin. An example here is Bohmian Mechanics. In this theory, it is not the case that the hidden state of the two-particle system at the time of preparation together with the choice of the direction of spin measurement uniquely determine the outcome of that measurement; moreover, the hidden state of the two-particle system will change between the time of preparation and the time of measurement. That hidden state, to recall, is the true positions of the two particles; since these particles need to move from the source to the detector, $\lambda_{t_P,1,2}$ will be different from $\lambda_{t_M, 1, 2}$, where $t_M$ is the time of measurement.%
\footnote{\label{fn:tM} There are several subtleties related to $t_M$ that will not matter for our discussion in the main text. First, it will typically not be the case that both measurements happen at exactly the same time (in a given reference frame). To avoid ambiguity, we can define $t_M$ to be the time of the first of the two measurements (or of the last of them). Second, in different reference frames the temporal relation between the two measurements will be different. However, now we are working with a fixed reference frame.}
Moreover, even though $\lambda_{t_M, 1, 2}$ is uniquely determined by $\lambda_{t_P}$ (because Bohmian mechanics is deterministic), it is not uniquely determined by $\lambda_{t_P,1,2}$ or even by $\lambda_{ \Sigma_{R_A, t_P} \cup \Sigma_{R_B, t_P}}$ (because Bohmian mechanics is not locally deterministic).%
\footnote{I consider here the version of Bohmian mechanics according to which the quantum state is not a beable but has a lawlike status because the version including the quantum state as a beable is not locally determinate (see section \ref{sec:locally-determinate}). The failure of local determinism is due to the fact that in Bohmian mechanics the motion of a given particle can be influenced by the positions of arbitrarily remote particles.}
Therefore, by our Bridge Principle Between Determinism and 0-1 Probability (Definition \ref{def:bridge}), in Bohmian mechanics it is \textit{not} the case that probabilities $P(A | a, \lambda_{t_P,1,2})$ and $P(B | b, \lambda_{t_P,1,2})$ have values $0$ or $1$. This means that Bohmian mechanics is not outcome deterministic \textit{despite being deterministic}.

In general, $P(A | a, \lambda_{t_P,1,2})$ and $P(B | b, \lambda_{t_P,1,2})$ having values $0$ or $1$ in accordance with $T$ is not guaranteed by $T$ being deterministic. If $T$ is locally deterministic, then $\lambda_{t_M, 1, 2}$ is uniquely determined by $\lambda_{ \Sigma_{R_A, t_P} \cup \Sigma_{R_B, t_P}}$ but still not necessarily by $\lambda_{t_P,1,2}$.%
\footnote{In the superdeterministic models LPW$_{\text{f}}$, C-LPW$_{\text{f}}$, LPW$_{\text{p}}$ and C-LPW$_{\text{p}}$ developed by \citet{ciepielewski-superdeterminism}, which are locally deterministic, $\lambda_{t_M, 1, 2}$ \textit{is} uniquely determined by $\lambda_{t_P,1,2}$. However, this is not a feature of all locally deterministic theories. The mentioned models possess this feature because, in these models, any point $(\vec{x}, t)$ is associated with a copy of the entire state of the universe at $t$, which evolves in accordance with the dynamics of Bohmian mechanics. Since the latter is deterministic, the laws of these models together with the state at any single spacetime point determine the evolution of the entire universe (under an additional assumption that the copies of the entire state of the universe are identical at every point). 
}

Let us use ``Factorizability$_{t_P, 1, 2}$'' and ``Settings Independence$_{t_P, 1, 2}$'' to denote the conditions of Factorizability and Settings Independence introduced in section \ref{sec:assumptions-Bell-theorem} in which ``$\lambda$'' is substituted with ``$\lambda_{t_P, 1, 2}$''. Settings Independence$_{t_P, 1, 2}$ seems very plausible: its violation would mean that the state of the measured system is correlated with the choices of measurement settings, so for some states of the system we cannot make or at least are less likely to make some types of measurements on it. If $\lambda_{t_P, 1, 2}$ lies in the causal past of $R_a$ and $R_b$, it \textit{can} influence the events that happen in these regions; however, the point is that there are so many \textit{other} events in the causal past of the choices of settings that the influence of $\lambda_{t_P, 1, 2}$ becomes insignificant (unless there exists some specific causal linkage between the preparation of the system and the choice of settings, which seems unlikely; cf. \citealt{shimony-local-beables}). It is also possible (at least in principle) to delay the preparation of the system so that it is spacelike related to the choices of measurement settings (see, e.g., \citealt[Fig.~6]{muller-placek-SI}), in which case nothing that happens in regions $R_a$ and $R_b$ can be influenced by $\lambda_{t_P, 1, 2}$ in a locally causal way. Some authors regard the violation of \textit{this} variant of Settings Independence (i.e., Settings Independence$_{t_P, 1, 2}$) as an option worth considering,%
\footnote{\label{fn:SI-violations} There are two variants of this view, depending on the direction of causal influence. According to superdeterminism, causation is directed towards the future (i.e., either $\lambda_{t_P, 1, 2}$ influences the settings, or something in their common past is responsible for the correlation between them). According to retrocausal approaches, causal influence travels into the past from settings to $\lambda_{t_P, 1, 2}$. Superdeterminism is considered, for example, by \citet{hossenfelder-palmer-superdet} and \citet{ciepielewski-superdeterminism}; retrocausality is considered, for example, by \citet{price-route-realism} and \citet{adlam-retrocausality} and reviewed by \citet{wharton-argaman-locally-mediated} and \citet{sep-qm-retrocausality}.} 
but for now let us assume that it holds. Then, from Bell's theorem it follows that there is no HVT that reproduces the predictions of QM and satisfies Factorizability$_{t_P, 1, 2}$. However, this does not answer the question of whether there exists an HVT that reproduces the predictions of QM and satisfies Factorizability with a different choice of $\lambda$ (i.e., encompassing more than the measured system but confined to the past light cones of $R_A$ and $R_B$, so that it remains relevant for local explanation of quantum correlations). Moreover, it is not clear how Factorizability$_{t_P, 1, 2}$ could be derived from Local Causality. The last problem is the most important one because such a derivation is indispensable for establishing \refbc{bell} on the basis of Bell's theorem with $\lambda = \lambda_{t_P, 1, 2}$.

At this point, one could respond that the proof of Bell's theorem works in the same way, no matter what $\lambda$ exactly is. Therefore, even if Factorizability$_{t_P, 1, 2}$ does not follow from Local Causality, we can find another variant of Factorizability with a different $\lambda$ that does follow from Local Causality, and run the proof of Bell's theorem with this $\lambda$ to establish \refbc{bell}. However, we will see that for some choices of $\lambda$, Settings Independence becomes less plausible, thereby undermining the entire reasoning. In particular, it will be shown that for $\lambda$ that encompasses full thick slices of the past light cones of measurements, any locally deterministic theory violates Settings Independence.

\section{Hidden state as a state of thick slices of the past light cones of measurements}\label{sec:state-of-region}

The second option for the choice of $\lambda$ is to take it to be the state of thick slices of the past light cones of measurements. This option was favoured by late Bell, and it is analysed in this section.

\subsection{Late Bell's concept of Local Causality}\label{subsec:state-of-region-Bell}

\begin{figure}[H]
\includegraphics[scale=0.56]{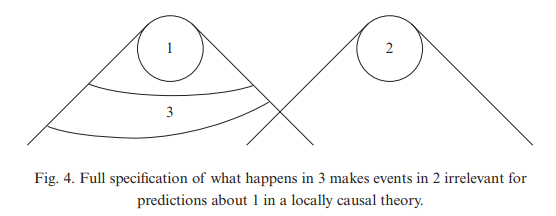}\includegraphics[scale=0.56]{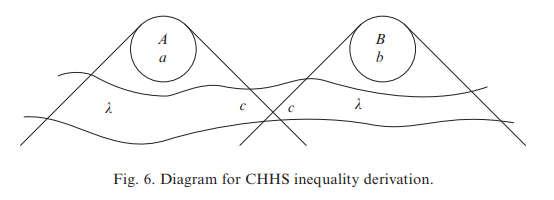}
\caption{Bell's (\citeyear[pp. 240 and 242]{Bell1990}) depiction of regions appearing in his formulation of Local Causality (left) and the derivation of the Bell-CHSH inequality (right).}\label{fig:bell-1990}
\end{figure}

\citet[p.~53]{Bell1976} introduces the concept of Local Causality as the generalisation of the concept of local determinism that is also applicable to indeterministic theories. In his last paper, \citet[p.~239]{Bell1990} formulates this principle as follows:
\begin{quote}
The direct causes (and effects) of events are near by, and even
the indirect causes (and effects) are no further away than
permitted by the velocity of light.
\end{quote}
He then provides a more precise formulation of it (\citeyear[pp.~239--240]{Bell1990}):
\begin{quote}
A theory will be said to be locally causal if the probabilities
attached to values of local beables in a space-time region 1 are unaltered by specification of values of local beables in a space-like
separated region 2, when what happens in the backward light cone
of 1 is already sufficiently specified, for example by a full specification of local beables in a space-time region 3 (Fig. 4 [which is on the left in my Fig. \ref{fig:bell-1990}]).
\end{quote}

\noindent
To this formulation, Bell adds the following commentary \citep[p.~240]{Bell1990}:
\begin{quote}
It is important that region 3 completely shields off from 1 the overlap
of the backward light cones of 1 and 2. And it is important that events
in 3 be specified completely. Otherwise the traces in region 2 of causes
of events in 1 could well supplement whatever else was being used for
calculating probabilities about 1. The hypothesis is that any such information about 2 becomes redundant when 3 is specified completely.
\end{quote}
This is in striking disagreement with the choice of $\lambda$ as $\lambda_{t_P, 1, 2}$, which lies in the overlap of the past light cones of $R_A$ and $R_B$, that is, Bell's ``overlap of the backward light cones of 1 and 2'' (cf. Fig. \ref{fig:Bell-exp}).%
\footnote{That Bell's formulation of his theorem and its assumptions changed over time has been recognised in the literature (see, e.g., \citealt{wiseman-bell-2014} and \citealt{brown-timpson-nonlocality}). However, the difference that is most often pointed out is that early Bell was concerned only with local HVTs that are ``deterministic'' in the sense of predicting the outcomes of spin measurements uniquely (and, relatedly, he considered only the quantum states that lead to the prediction of perfect (anti-)correlations, since these are the states relevant for the EPR argument), whereas later he wanted to also encompass local HVTs that give only probabilistic predictions. 
Here, I have assumed from the start that Bell's theorem should be applicable to theories of both types; the differences between early and late Bell that I discuss are only those regarding the localisation of $\lambda$.}

Here is a proposal for how to formulate this condition mathematically (cf. a similar formulation in \citealt[p.~277]{norsen-bell-jarrett-2009}):

\begin{definition}\label{def:LC-fg}
Local Causality$_{C(\Sigma, \Sigma)}$, fine-grained version: For any bounded region $R$ and any bounded region $R'$ that is space-like related to it, 
for any $t,t' < R$ that lie above the intersection of the past light cones of $R$ and $R'$ and such that $t'<t$, and for any $\lambda_R$,
$\lambda_{R'}$ and $\lambda_{C(\Sigma_{R, t}, \Sigma_{R, t'})}$,%
\footnote{The quantification over $\lambda_R$ is the quantification over all hidden states of a given HVT that can be ascribed to region $R$ (other quantifiers are understood analogously).
To make this fully rigorous, we should proceed as follows: define function $\boldsymbol{\lambda}$ that to any region $R$ assigns the set of all hidden states of a given HVT that can be ascribed to this region, and replace ``for all $\lambda_R$'' with ``for all $\lambda \in \boldsymbol{\lambda} (R)$''.
}
\begin{equation}\label{eq:local-causality}
P(\lambda_R |  \lambda_{C(\Sigma_{R, t}, \Sigma_{R, t'})}, \lambda_{R'})  =
P(\lambda_R |  \lambda_{C(\Sigma_{R, t}, \Sigma_{R, t'})}).
\end{equation}
\end{definition}

\begin{figure}
\centering
\includegraphics[width=\textwidth]{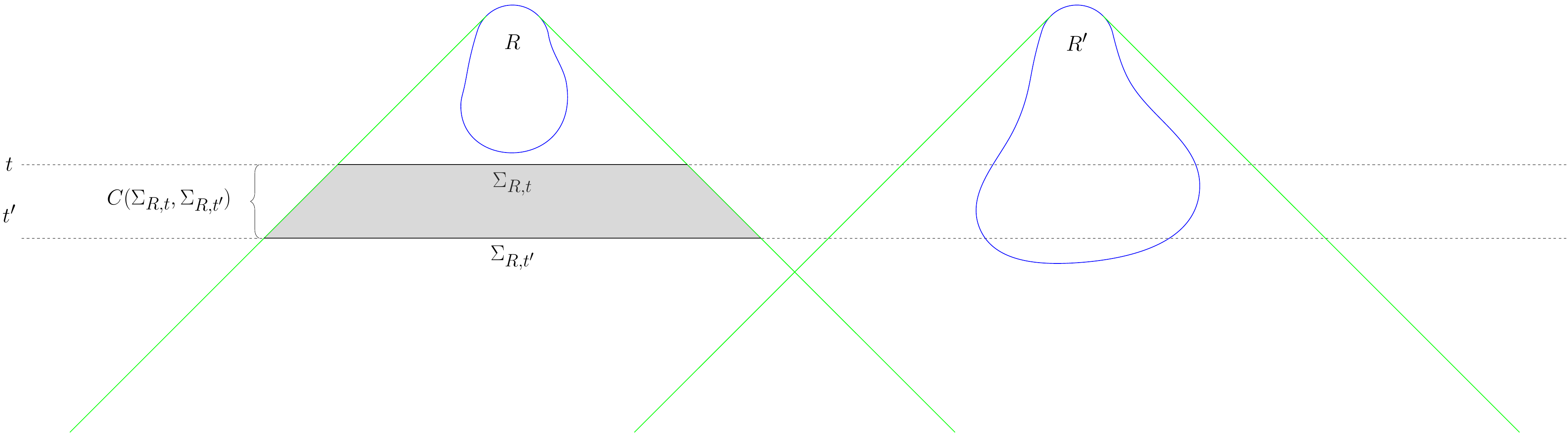}
\caption{Relations between the regions mentioned in the formulation of Local Causality$_{C(\Sigma, \Sigma)}$.
In the proof of Factorizability$_{C(\Sigma, \Sigma)}$, we substitute for $R$ either region of the choice of settings (i.e., $R_a$/$R_b$) or the region of measurement together with the region of the choice of settings (i.e., $R_A \cup R_a$/$R_B \cup R_b$). For details see Appendix \ref{app:Fact-from-LC}. }\label{fig:LC}
\end{figure}

\begin{figure}
\includegraphics[width=\textwidth]{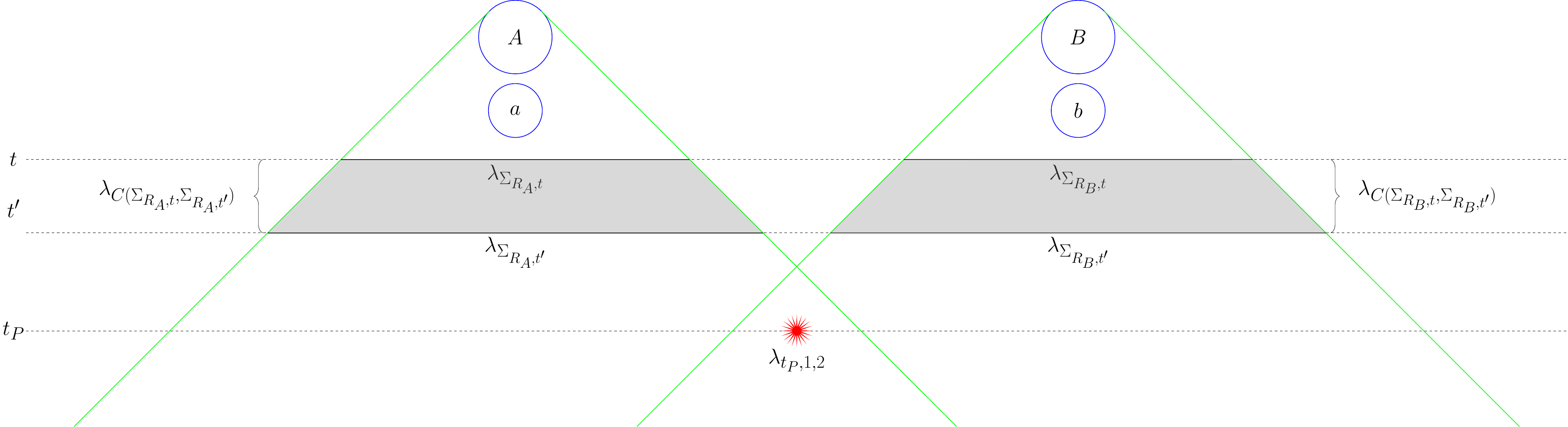}
\caption{Comparison of two different choices of $\lambda$ in Bell's theorem, $\lambda_{t_P, 1, 2}$ vs. $\lambda_{C(\Sigma, \Sigma)}$.}\label{fig:Bell-exp}
\end{figure}

The respective regions are depicted in Fig. \ref{fig:LC}. Regions $R$ and $R'$ are assumed to be bounded but not necessarily connected, which will be exploited later (see Appendix \ref{app:Fact-from-LC}). Notice that in concrete calculations this principle can be used in two ways: either ``from left to right'' (i.e., we can remove conditionalization on what happens in $R'$) or ``from right to left'' (i.e., we can harmlessly add conditionalization on what happens in $R'$).

The main difference with Bell's (1990) formulation is that the above formulation uses a fixed slicing in order to define the region that is supposed to screen off the region $R$ from the region $R'$. This might seem to be in tension with relativity theory, which rejects the existence of a distinguished slicing; however, if we add the requirement that Local Causality$_{C(\Sigma, \Sigma)}$ should be satisfied in any slicing, then the tension disappears. We could also consider regions of more ``wavy'' shapes, as depicted in Fig. \ref{fig:bell-1990}, but I prefer using $\Sigma_{R, t}$ to define these regions because this will make it easier to relate Local Causality$_{C(\Sigma, \Sigma)}$ to locally deterministic theories.

The above version of Local Causality$_{C(\Sigma, \Sigma)}$ is not quite suitable for our purposes because all arguments of the probability function are complete specifications of hidden variables (beables) in respective regions, whereas we are interested in representing more coarse-grained specifications. For example, we are not interested in all differences between possible specifications of beables in the region $R_a$ (i.e., possible $\lambda_{R_a}$'s), so we should consider their equivalence classes (each of which corresponds to one choice of measurement setting in region $R_a$) instead of single such specifications.  In order to capture this, we will introduce a further piece of notation. Let $d_a$ be a (coarse-grained) description of an event that can happen in a region $R_a$ which can be formulated within a given HVT. Then, by $\overline{\lambda_{R_a; d_a}}$ we denote the set of all and only those $\lambda_{R_a}$'s in which $d_a$ is true. 
Notice that since we use $d_a$ to define $\overline{\lambda_{R_a; d_a}}$, not all sets of states count as coarse-grained states but only those definable within the theory.
This leads to another version of Local Causality:

\begin{definition}\label{def:LC-cg}
Local Causality$_{C(\Sigma, \Sigma)}$, coarse-grained version: 
For any bounded region $R$ and any bounded region $R'$ that is space-like related to it, 
for any $t,t' < R$ that lie above the intersection of the past light cones of $R$ and $R'$ and such that $t'<t$, for any $\lambda_R$, $\lambda_{R'}$ and $\lambda_{C(\Sigma_{R, t}, \Sigma_{R, t'})}$,  and for any coarse-grainings $\overline{\lambda_{R, d_{R}}}$ and $\overline{\lambda_{R', d_{R'}}}$,
\begin{equation}\label{eq:local-causality-coarse-grained}
P(\overline{\lambda_{R, d_{R}}} |  \lambda_{C(\Sigma_{R, t}, \Sigma_{R, t'})}, \overline{\lambda_{R', d_{R'}}})  =
P(\overline{\lambda_{R, d_{R}}} | \lambda_{C(\Sigma_{R, t}, \Sigma_{R, t'})}).
\end{equation}
\end{definition}

Notice that all complete specifications of beables are now replaced by coarse-grained ones, except for the region $C(\Sigma_{R, t}, \Sigma_{R, t'})$, where we want to retain a complete specification of beables, in agreement with Bell. For any coarse-grained state, what matters are not the probabilities associated with complete states that fall under it, but only the probabilities associated with the entire equivalence class. In our description of experiments, instead of the above complex notation for coarse-grained states, we will use the values of chosen settings and the values of obtained outcomes, which will be denoted, as previously, by $a, b, A, B$ etc.

It can be shown (see Appendix \ref{app:Fact-from-LC}) that Local Causality$_{C(\Sigma, \Sigma)}$ (in the coarse-grained version) entails Factorizability$_{C(\Sigma, \Sigma)}$, that is, 
\begin{equation}
\begin{split}
P(A, B | a, b, \lambda_{C(\Sigma, \Sigma)}) =
P(A | a, \lambda_{C(\Sigma, \Sigma)}) 
P(B | b, \lambda_{C(\Sigma, \Sigma)}),
\end{split}
\end{equation}
where $\lambda_{C(\Sigma, \Sigma)}$ denotes the pair consisting of $\lambda_{C(\Sigma_{R_A, t}, \Sigma_{R_A, t'})}$ and $\lambda_{C(\Sigma_{R_B, t}, \Sigma_{R_B, t'})}$.

We can also formulate a variant of Settings Independence for our choice of hidden variables, denoted by Settings Independence$_{C(\Sigma, \Sigma)}$:
\begin{equation}
P ( \lambda_{C(\Sigma, \Sigma)} | a, b) =
P  ( \lambda_{C(\Sigma, \Sigma)}).
\end{equation}
This completes our formulation of the assumptions of Bell's theorem for hidden states encompassing entire thick slices of the past light cones of measurement regions $R_A$ and $R_B$.%

\subsection{Locally deterministic theories vs. Local Causality$_{C(\Sigma, \Sigma)}$ and Settings Independence$_{C(\Sigma, \Sigma)}$}\label{subsec:state-of-region-loc-det}

One can show that any locally deterministic HVT satisfies Local Causality$_{C(\Sigma, \Sigma)}$ but violates Settings Independence$_{C(\Sigma, \Sigma)}$. This is the content of the following two theorems (for proofs see Appendix \ref{app:loc-det-LC-SI}):

\begin{theorem}\label{thm:local-det-LocalCausality}
If $T$ is a locally deterministic HVT, then it satisfies Local Causality$_{C(\Sigma, \Sigma)}$.
\end{theorem}

\begin{theorem}\label{thm:local-det-violationSettInd}
If $T$ is a locally deterministic HVT, then it violates Settings Independence$_{C(\Sigma, \Sigma)}$.
\end{theorem}

In fact, all locally deterministic theories violate even a much weaker condition, which might be called Settings Compatibility$_{C(\Sigma, \Sigma)}$ and which states that any $\lambda_{C(\Sigma, \Sigma)}$ is compatible with any choice of settings in $R_a \cup R_b$. In locally deterministic theories, any $\lambda_{C(\Sigma, \Sigma)}$ is compatible with \textit{exactly one} choice of such settings.

One can object to the above reasoning by pointing out that it relies on the following two additional implicit assumptions:
\begin{quote}
Physical Nature of Human Choices: Human choices are physical processes.
\end{quote}
\begin{quote}
Universality of Laws: The laws of $T$ apply to all physical processes.
\end{quote}
If the Universality of Laws holds, then all physical processes must conform to the laws of $T$.
If, in addition, the Physical Nature of Human Choices holds, then, in particular, the choice of measurement settings by human experimenters must conform to the laws of $T$. This is precisely what we assumed implicitly in the proof of Theorem \ref{thm:local-det-violationSettInd} when we applied the laws of $T$ to the physical contents of regions $R_a$ and $R_b$. Therefore, our conclusion that any locally deterministic theory violates Settings Independence$_{C(\Sigma, \Sigma)}$ can be avoided by rejecting one of these assumptions.

If we reject the Universality of Laws (while retaining the Physical Nature of Human Choices), then a locally deterministic theory can be made consistent with Settings Independence$_{C(\Sigma, \Sigma)}$, provided that the domain of application of the laws of $T$ is appropriately restricted.%
\footnote{This seems to be somewhat similar to the approach of \citet{muller-placek-SI}, with their distinction between nature-induced indeterminism and agent-induced indeterminism. However, this is not quite the same because they do not talk about laws.}
More specifically, we need to assume that $T$ does not apply to some types of physical processes that are involved in choices of measurement settings. Importantly, rejecting the Universality of Laws does not need to mean that we just create a gap in the natural order to make room for humans' freedom as there might be other types of physical processes that are not subject to the laws of $T$ and are not in any way related to humans.

The drawback of this strategy is that by rejecting the Universality of Laws we also block our proof that any locally deterministic theory satisfies Local Causality$_{C(\Sigma, \Sigma)}$. Therefore, this strategy is not helpful for establishing \refbc{bell}, since---as mentioned above---Local Causality was intended to be a generalisation of the idea of locality as it appears in locally deterministic theories.

We can also reject the Physical Nature of Human Choices while retaining the Universality of Laws. The problem with this option is that in most actual realisations of Bell's Experiment, the measurement settings are not chosen by humans but by means of some physical processes (an exception is an experiment done by The BIG Bell Test Collaboration, see \citealt{big-bell-test}). This would mean that only experiments involving human choices are actual tests of Local Causality and all others are not, which might seem surprising given that both types of experiments yield the same results (namely, the violation of the Bell-CHSH inequality in agreement with QM predictions).

Some statements made by Bell himself might be interpreted as the rejection of either the Physical Nature of Human Choices or the Universality of Laws (or both). For example, \citet[p.~61]{Bell1976} writes:
\begin{quote}
It has been assumed that the settings of instruments are in some sense free
variables---say at the whim of experimenters---or in any case not determined in the overlap of the backward light cones. 
\end{quote}
In a subsequent paper, he elaborates on this fragment as follows \citep[p.~101]{Bell1977}:
\begin{quote}
A respectable class of theories, including contemporary quantum theory
as it is practised, have `free' `external' variables in addition to those internal
to and conditioned by the theory. These variables are typically external
fields or sources. They are invoked to represent experimental conditions.
They also provide a point of leverage for `free willed experimenters', if
reference to such hypothetical metaphysical entities is permitted. I am
inclined to pay particular attention to theories of this kind [...] 
\end{quote}
If the mentioned external parameters are irreducibly external in the sense that they cannot be explained by applying the theory to a larger system, then this indeed looks like a violation of the Physical Nature of Human Choices or the Universality of Laws (or both).

However, soon after, Bell weakens the assumption that the experimental settings are free to the assumption that they are effectively or sufficiently free ``for the purpose at hand''. He invokes the example of a deterministic number generator (\citealt[pp.~102--103]{Bell1977}; this is also the line that he takes in his last paper---see \citealt[p.~244]{Bell1990}):
\begin{quote}
A particular output is the result
of combining so many factors, of such a lengthy and complicated dynamical chain, that it is quite extraordinarily sensitive to minute variations of any
one of many initial conditions. It is the familiar paradox of classical
statistical mechanics that such exquisite sensitivity to initial conditions is
practically equivalent to complete forgetfulness of them. To illustrate the
point, suppose that the choice between two possible outputs, corresponding to $a$ and $a'$ depended on the oddness or evenness of the digit in the
millionth decimal place of some input variable. Then fixing $a$ or $a'$ indeed
fixes something about the input---i.e., whether the millionth digit is odd
or even. But this peculiar piece of information is unlikely to be the vital
piece for any distinctively different purpose [...]
\end{quote}
Bell does not specify which particular ``distinctively different purpose'' is relevant for establishing \refbc{bell}. If we take it to be ``rendering the far-away outcome $B$ and setting $b$ redundant for the task of determining the probabilities of obtaining $A$'' \citep[p.~445]{seevinck-uffink-bell}, then we obtain the approach that will be discussed in section \ref{sec:incomplete-sufficient}.

\subsection{Further discussion of the violation of Settings Independence$_{C(\Sigma, \Sigma)}$ in locally deterministic theories}\label{subsec:state-of-region-SettInd}

The fact that all locally deterministic theories violate Settings Independence$_{C(\Sigma, \Sigma)}$ (and even a much weaker condition of Settings Compatibility$_{C(\Sigma, \Sigma)}$) might be worrying because Settings Independence is sometimes argued to be indispensable for experimental practice. If these arguments were able to establish that Settings Independence$_{C(\Sigma, \Sigma)}$ holds, then we could exclude all locally deterministic theories without using Bell's theorem, merely on the basis of Theorem \ref{thm:local-det-violationSettInd}. However, under closer scrutiny, typical arguments against Settings Independence turn out to be much less convincing in the case of Settings Independence$_{C(\Sigma, \Sigma)}$ than in the case of Settings Independence$_{t_P, 1, 2}$ or, more generally, Settings Independence$_{\lambda_\text{system}}$. 

Let us briefly look at four such arguments. The first argument is that Settings Independence is indispensable for the procedure of randomization in empirical sciences such as medicine. \citet{scholarpedia-bell} give the following example:
\begin{quote}
[...] if you are performing a drug versus placebo clinical trial, then you have to select some group of patients to get the drug and some group of patients to get the placebo. The conclusions drawn from the study will necessarily depend on the assumption that the method of selection is independent of whatever characteristics those patients might have that might influence how they react to the drug.
\end{quote}
However, what they say clearly applies to Settings Independence$_{\lambda_\text{system}}$ rather than Settings Independence$_{C(\Sigma, \Sigma)}$. This is visible in another fragment of \citet{scholarpedia-bell} (emphasis mine): 
\begin{quote}
[...] in practice, one assesses the applicability of [Settings Independence] to a given experiment by examining the care with which the experimental design precludes any non-conspiratorial dependencies \textit{between the preparation of the systems and the settings of instruments}.
\end{quote}
Moreover, medicine, biology and social sciences are usually or even always interested in coarse-grained variables, not in full specifications of the physical state of the systems they are investigating, whereas Settings Independence$_{C(\Sigma, \Sigma)}$ asserts the dependence of the choice of settings on the \emph{full specification} of the state in $C(\Sigma, \Sigma)$.

The second argument relies on the concept of a causal linkage (\citealt{shimony-local-beables}; cited from \citealt[p.~168]{shimony-naturalistic}): 
\begin{quote}
Bell can, of course, reply that we do not know that the distribution
of emissions [$\lambda$] is insensitive to the values of $a$ and $b$, or for that matter
that there are no causal links between the act of selecting $a$ and that of
selecting $b$. After all, the backward light cones of those two acts do eventually overlap, and one can imagine one region which controls the decision of the two experimenters who chose $a$ and $b$. We cannot deny such
a possibility. But we feel that it is wrong on methodological grounds to
worry seriously about it if no specific causal linkage is proposed. 
\end{quote}
Again, these authors clearly have in mind Settings Independence$_{t_P, 1, 2}$ and their argument cannot be extended to Settings Independence$_{C(\Sigma, \Sigma)}$. In the latter case, the dependence of the choice of settings on $\lambda_{C(\Sigma, \Sigma) }$ is guaranteed by the laws of the locally deterministic theory, irrespective of whether this relationship can also be captured in terms of ``causal linkages''. 

The third argument, which also has its roots in the paper by \citet{shimony-local-beables}, is that rejecting Settings Independence amounts to abandoning ``in advance the whole enterprise of discovering the laws of nature by experimentation'' (cited from \citealt[p.~168]{shimony-naturalistic}). The issue of discovering physical laws is more relevant for the discussion of Settings Independence$_{C(\Sigma, \Sigma)}$ because we test hypotheses about such laws by preparing a system in some initial state and checking whether its evolution agrees with that prescribed by these hypothetical laws, so the full specification of a physical state (and not merely a coarse-grained one) becomes important. 
In order to learn about laws of nature from such experiments, it should be possible to measure any aspect of the final state of the system, no matter what its initial state is (and perhaps also our likelihood of measuring this particular aspect should not depend on the initial state). If this condition were violated, then it could happen that for any initial state we cannot measure exactly those aspects of the final state which disagree with our hypothetical laws, thereby making our tests unreliable. However, as before, testing laws requires only Settings Independence$_{\lambda_\text{system}}$ and not Settings Independence$_{C(\Sigma, \Sigma)}$.%
\footnote{There are other problems with testing HVTs in this way: for example, for some such theories we might not be able to know the complete initial and final state of the system sufficiently well to test the predictions derived from the hypothetical laws. However, this problem has nothing to do with Settings Independence.}

The fourth argument is that violation of Settings Independence ``requires an atypical fine-tuning of the initial state of the universe'' (see, e.g., \citealt[p.~10]{baas-bihan-superdeterminism}). Again, fine-tuning might be needed in the case of violations of Settings Independence$_{t_P, 1, 2}$ (although one can construct a theory in which the correlations between settings and $\lambda_{t_P, 1, 2}$ are lawlike and, as such, independent of the initial state; see, e.g., \citealt{ciepielewski-superdeterminism}) but not in the case of violations of Settings Independence$_{C(\Sigma, \Sigma)}$. This is reflected in the fact that our proof of Theorem \ref{thm:local-det-violationSettInd} does not make any assumptions about the initial state of the universe.

\section{Attempts at improvement}\label{sec:improve}

The results of sections \ref{sec:state-of-system} and \ref{sec:state-of-region} pose a serious threat to \refbc{bell}. We have considered two possible choices of $\lambda$, neither of which leads straightforwardly to any conclusion about Local Causality$_{C(\Sigma, \Sigma)}$:
\begin{itemize}
\item If we consider $\lambda = \lambda_{t_P, 1, 2}$, then we do not take into account other possible local explanations of correlations between measurement outcomes. Moreover, the relationship between Factorizability$_{t_P, 1, 2}$ and Local Causality$_{C(\Sigma, \Sigma)}$ is not clear. 
\item The choice of $\lambda = \lambda_{C(\Sigma, \Sigma)}$ might seem better because Factorizability$_{C(\Sigma, \Sigma)}$ follows from Local Causality$_{C(\Sigma, \Sigma)}$, but then we cannot run the proof of Bell's theorem for locally deterministic theories (which form an important subclass of locally causal theories) because they violate Settings Independence$_{C(\Sigma, \Sigma)}$ (at least if we assume the Universality of Laws and the Physical Nature of Human Choices).
\end{itemize}

As far as I can see, there are four possible ways to try to recover \refbc{bell} in the face of these difficulties: 
\begin{itemize}
\item take Factorizability$_{t_P, 1, 2}$ rather than Local Causality$_{C(\Sigma, \Sigma)}$ as the proper formulation of local causality;
\item instead of complete states $\lambda_{C(\Sigma, \Sigma)}$,
use states $\overline{\lambda}$ specified incompletely (so that Settings Independence$_{\overline{\lambda}}$ is satisfied) but sufficiently (so that Factorizability$_{\overline{\lambda}}$ is satisfied);
\item derive Factorizability$_{t_P, 1, 2}$ from Local Causality$_{C(\Sigma, \Sigma)}$, perhaps using some additional but plausible assumptions;
\item consider some different localisations of hidden states.
\end{itemize}
The first three options will be analysed in the following subsections. Notice that the first and third options regard $\lambda_{t_P, 1,2}$ as the proper choice of $\lambda$ for Bell's theorem; the second option uses $\lambda_{C(\Sigma, \Sigma)}$ but takes a coarse-graining of it; and the last option looks for yet another choice of $\lambda$.

\subsection{Local causality is Factorizability$_{t_P, 1, 2}$}

Why can we not be content with showing that Factorizability$_{t_P, 1, 2}$ is false? Can this principle (instead of Local Causality$_{C(\Sigma, \Sigma)}$) be regarded as the proper formulation of the intuitive idea of local causality? \citet[p.~243]{Bell1990} was clearly unwilling to accept this, since he writes: 
\begin{quote}
Very often such factorizability is taken as the starting point
of the analysis. Here we have preferred to see it not as the formulation
of `local causality', but as a consequence thereof.
\end{quote}
He does not provide an in-depth justification for this conceptual choice, but I believe that he had good reasons for it. 

Bell's focus on Local Causality$_{C(\Sigma, \Sigma)}$ seems to be in line with his broader views about physics. Factorizability$_{t_P, 1, 2}$ is a condition that is applicable only to situations of a very special type---namely, the instances of Bell's Experiment (and other appropriately similar experiments). 
This is because, in its formulation, Factorizability$_{t_P, 1, 2}$ involves concepts such as measurement settings and measurement outcomes, and presupposes a certain structure of the experiment.
In contrast, Local Causality$_{C(\Sigma, \Sigma)}$ is a very general physical principle---in fact, it is applicable to any physical situation. Moreover, Local Causality$_{C(\Sigma, \Sigma)}$ does not involve in its formulation the concept of measurement, and one needs to remember that for Bell, the use of the concept of measurement in the formulation of the Copenhagen interpretation of QM was its main disadvantage (cf. \citealt{Bell1989}). For this reason, Local Causality$_{C(\Sigma, \Sigma)}$ seems more relevant for the foundations of physics than Factorizability$_{t_P, 1, 2}$.

Another possible objection to Factorizability$_{t_P, 1, 2}$ is that it relies not only on the idea of locality but also on the common cause principle (see, e.g. \citealt[pp.~8--9]{Barandes2024}). This principle, which was formulated by Reichenbach, asserts that whenever events $A$ and $B$ are correlated, that is, 
\begin{equation}
P(A, B) \neq P(A)  P(B),
\end{equation}
then there exists another event $C$ such that
\begin{equation}\label{eq:common-cause}
P(A, B| C) = P(A | C)  P(B | C)
\end{equation}
(for the full statement, see \citealt{sep-common-cause}). 
In contradistinction to Factorizability$_{t_P, 1, 2}$, Local Causality$_{C(\Sigma, \Sigma)}$ does not involve the concept of common cause in any way. In Definitions \ref{def:LC-fg} and \ref{def:LC-cg}, $\lambda_{C(\Sigma_{R, t}, \Sigma_{R, t'})}$ is not supposed to be a common cause of some two events but a cause of events in region $R$. 

The exact justification that Definitions \ref{def:LC-fg} and \ref{def:LC-cg} correctly capture the intuitive idea of local causality is another matter. One possible justification relies on analogical reasoning. In the case of locally deterministic theories, state $\lambda_{C(\Sigma_{R, t}, \Sigma_{R, t'})}$ uniquely determines state $\lambda_R$, so information about state in any other region is redundant for $\lambda_R$ given $\lambda_{C(\Sigma_{R, t}, \Sigma_{R, t'})}$. 
Now, Local Causality$_{C(\Sigma, \Sigma)}$ is supposed to be the generalisation of this idea to theories that are locally causal (in an intuitive sense) but not necessarily locally deterministic.
If we replace ``unique determination of state'' with ``unique determination of the \emph{probability of} state'', we obtain Definitions \ref{def:LC-fg} and \ref{def:LC-cg} (but now we must remember that information about events lying within the light cone of $R$ after $t$ might \emph{not} be redundant).

\subsection{The idea of an incomplete but sufficient specification of hidden states}\label{sec:incomplete-sufficient}

The idea of considering coarse-grained versions of $\lambda_{C(\Sigma, \Sigma)}$ arises from Bell's expression ``what happens in the backward light cone of 1 is already sufficiently specified, for example by a full specification of local beables in a space-time region 3'' (cf. the full quote in section \ref{subsec:state-of-region-Bell}). We obtained the result that any locally deterministic theory violates Settings Independence$_{C(\Sigma, \Sigma)}$ because we have taken $\lambda$ to be the \textit{complete} specification of the physical contents of the region $C(\Sigma, \Sigma)$. However, Bell might be understood as saying only that $\lambda$ should be a \textit{sufficient} specification; it might be, for example, a complete specification, but perhaps some incomplete specifications are sufficient as well. This idea appears in Norsen's (\citeyear[p.~283]{norsen-bell-jarrett-2009} and \citeyear[p.~1266]{norsen-bell-2011}) analysis of Bell and was taken up by \citet{seevinck-uffink-bell}. The latter authors propose a precisification of the concept of sufficiency. The crucial question is, of course, for what purpose should the specification of $\lambda$ be sufficient? Their answer is that it should be sufficient ``for the purpose of rendering the far-away outcome $B$ and setting $b$ redundant for the task of determining the probabilities of obtaining $A$'' \citep[p.~445]{seevinck-uffink-bell}. This means that we should consider $\overline{\lambda}$'s that satisfy the following condition (cf. \citealt[p.~445]{seevinck-uffink-bell}):
\begin{equation}\label{eq:sufficient-lambda}
P(A | a, b, B, \overline{\lambda}) = P(A | a, \overline{\lambda}) \text{ and } P(B| a, b, A, \overline{\lambda}) = P(B | b, \overline{\lambda}).
\end{equation}

Clearly, something less than the full $\lambda_{C(\Sigma, \Sigma)}$ might satisfy this condition. For example, in a locally deterministic theory, 
\begin{equation}
\overline{\lambda} = \biggl\{ \{ \lambda_{C(\Sigma_{R_A, t}, \Sigma_{R_A, t'})}, \lambda_{C(\Sigma_{R_B, t}, \Sigma_{R_B, t'})} \},
\{ \lambda'_{C(\Sigma_{R_A, t}, \Sigma_{R_A, t'})} , \lambda'_{C(\Sigma_{R_B, t}, \Sigma_{R_B, t'})} \} \biggr\}
\end{equation} 
where $\lambda_{C(\Sigma_{R_A, t}, \Sigma_{R_A, t'})}$ is consistent with $a$,
$\lambda_{C(\Sigma_{R_B, t}, \Sigma_{R_B, t'})}$ is consistent with $b$,
$\lambda'_{C(\Sigma_{R_A, t}, \Sigma_{R_A, t'})} $ is consistent with $a' \neq a$
and $\lambda'_{C(\Sigma_{R_B, t}, \Sigma_{R_B, t'})}$ is consistent with $b \neq b'$,
is sufficient in this sense. This example also shows that incomplete specifications of the physical contents of some region $R$ can be constructed by taking classes of $\lambda_R$'s instead of separate such states.%
\footnote{This distinguishes my approach from that of \citet{seevinck-uffink-bell}, who prefer to talk in terms of variables instead of in terms of classes of complete states. However, any statements made within the former approach can be translated into statements within the latter approach because any specification of some variables of a theory $T$ in a region $R$ corresponds to a class of possible complete states of region $R$. Another difference is that Seevinck and Uffink regard $a$ and $b$ as labels of probability functions, while I keep the convention that they are arguments of a single probability function.} 
The hope is that we can impoverish our complete states (by replacing them with classes of such states) so that they will still be sufficient, but by omitting information we will lose the correlation between these states and the choice of settings. 

There are two problems with the above concept of sufficient specification. The first is that, in general, we do not even know whether the complete specification is sufficient! In a nonlocal theory, the complete specification of beables in the region $C(\Sigma, \Sigma)$ is not sufficient in the above sense---this is the whole point of the theory being nonlocal. Therefore, \textit{a fortiori}, we cannot exclude that no incomplete specification is sufficient. We need to formulate our question in the following way: Is it the case that for any HVT that satisfies Local Causality$_{C(\Sigma, \Sigma)}$, there exists a coarse-graining of $\Lambda$ (call it $\overline{\Lambda}$), such that the elements of $\overline{\Lambda}$ satisfy sufficiency condition \eqref{eq:sufficient-lambda} and Settings Independence$_{\overline{\lambda}}$? If so, then we can run Bell's theorem with this $\overline{\lambda}$ and show that any such HVT is inconsistent with the predictions of QM. 

The second problem with the proposal of \citet{seevinck-uffink-bell} is that their sufficiency condition is unnecessarily strong. This is because what we really want is to draw from Bell's theorem some conclusions about any HVT that satisfies Local Causality$_{C(\Sigma, \Sigma)}$, and for this purpose, it is sufficient that at least one version of Factorizability and Settings Independence (but with the same $\overline{\lambda} $!) is satisfied by this HVT. Therefore, I suggest that instead of \eqref{eq:sufficient-lambda} we should just use Factorizability$_{\overline{\lambda}}$. 

With these amendments, the reasoning that one can use to exclude all HVTs that satisfy Local Causality$_{C (\Sigma, \Sigma)}$ as inconsistent with empirically confirmed predictions of QM is as follows.
Assume that for any theory satisfying Local Causality$_{C (\Sigma, \Sigma)}$, there exists $\overline{\Lambda} \subsetneq \mathcal{P}(\Lambda) $ such that $ \bigcup \overline{\Lambda} = \Lambda$, the elements of $\overline{\Lambda}$ are pairwise disjoint, and both Factorizability$_{\overline{\lambda}}$ and Settings Independence$_{\overline{\lambda}}$ hold. Let $T$ be any HVT satisfying Local Causality$_{C (\Sigma, \Sigma)}$. Then, for this theory $T$, there exists $\overline{\Lambda}$ such that both Factorizability$_{\overline{\lambda}}$ and Settings Independence$_{\overline{\lambda}}$ are satisfied (by our assumption). However, Factorizability$_{\overline{\lambda}}$ and Settings Independence$_{\overline{\lambda}}$ entail Bell's inequality, which is violated. Therefore, $T$ cannot be true. Since $T$ was an arbitrary HVT that satisfies Local Causality$_{C (\Sigma, \Sigma)}$, we have shown in this way that Local Causality$_{C (\Sigma, \Sigma)}$ is false.

The problem with this reasoning is that it is not obvious that for every HVT satisfying Local Causality$_{C (\Sigma, \Sigma)}$ at least one such $\overline{\Lambda}$ exists. Could it be the case that whenever we coarse-grain our hidden states so that they satisfy Settings Independence$_{\overline{\lambda}}$, they also cease to satisfy Factorizability$_{\overline{\lambda}}$?
It seems that one might as well run the reasoning that is opposite to that in the previous paragraph: if an HVT reproduces the predictions of QM, then its coarse-grained states cannot satisfy both Settings Independence$_{\overline{\lambda}}$ and Factorizability$_{\overline{\lambda}}$. How can one break the impasse between these two lines of reasoning?%
\footnote{\citet{hofer-szabo-SU} argues that it is logically possible that the required coarse-graining exists. However, our question is different: does the required coarse-graining exist for all HVTs that satisfy Local Causality$_{C (\Sigma, \Sigma)}$ (or at least for a large class of them, which leaves outside only those HVTs satisfying Local Causality$_{C (\Sigma, \Sigma)}$ that are implausible for some independent reason)?}

One possible response has been proposed by \citet{scholarpedia-bell}. Although they do not use the terminology of incompleteness and sufficiency, their proposal is effectively very similar to the one described above. Instead of talking about coarse-grained versions of $\lambda_{C (\Sigma, \Sigma)}$, they consider $\lambda_{C (\Sigma, \Sigma)}$ as divided into several beables, some of which are irrelevant for the measurement outcome, while the others either influence the outcome only via their influence on measurement settings or in some other way (the latter correspond to our $\overline{\lambda}$). Their solution to our problem is that theories in which $\lambda_{C (\Sigma, \Sigma)}$ cannot be divided in this way are called ``conspiratory'' and are excluded from the scope of Bell's theorem. However, one might worry that in this approach, Bell's theorem would be inapplicable to an excessively large class of theories, some of which cannot be excluded on independent grounds. The crucial question is whether we have equally strong reasons for Settings Independence$_{\overline{\lambda}}$ as we had for Settings Independence$_{t_P, 1, 2}$. This is unclear because the arguments for the latter (see section \ref{subsec:state-of-region-SettInd}) concern $\lambda_{t_P, 1,2}$, which is the state of the measured system at the moment of preparation, whereas $\overline{\lambda}$ is a much more encompassing state with a different localisation in spacetime, so these arguments cannot be adapted straightforwardly to support Settings Independence$_{\overline{\lambda}}$. However, other equally strong arguments may exist;
one argument sketched by Bell (see the last quotation in section \ref{subsec:state-of-region-loc-det}) can be read as supporting Settings Independence$_{\overline{\lambda}}$.

\subsection{Attempts to derive Factorizability$_{t_P, 1, 2}$ from Local Causality$_{C(\Sigma, \Sigma)}$}\label{sec:derive-fact-from-lc}

The motivation for the next approach can be formulated as follows. Settings Independence$_{t_P, 1, 2}$ is plausible, so if we could derive Factorizability$_{t_P, 1, 2}$ from Local Causality$_{C(\Sigma, \Sigma)}$, then we could show that all locally causal theories are excluded as inconsistent with QM by the following reasoning:

\inferencel{Local Causality$_{C (\Sigma, \Sigma)}$ $\Rightarrow$ 
Factorizability$_{t_P, 1, 2}$ 
\and Settings Independence$_{t_P, 1, 2}$  
\and $\bigl($Factorizability$_{t_P, 1, 2}$ $\wedge$ Settings Independence$_{t_P, 1, 2} \bigr)$ $\Rightarrow$ Bell's inequality 
\and $\neg$ Bell's inequality}%
{$\neg$Local Causality$_{C (\Sigma, \Sigma)}$}

The problematic step in this reasoning is, of course, the first one. It is clear that Factorizability$_{t_P, 1, 2}$ does not follow logically from Local Causality$_{C (\Sigma, \Sigma)}$ because they involve different states (i.e., $\lambda_{t_P,1,2}$ and $\lambda_{C(\Sigma, \Sigma)}$, respectively) that are ascribed to different regions and are not definitionally related to each other. Therefore, if there is any chance of saving the first step, this would require some additional assumptions. Of course these additional assumptions should be independently plausible and have some physical meaning. On the other hand, the additional assumptions should not entail Factorizability$_{t_P, 1, 2}$ on their own, independently of Local Causality$_{C (\Sigma, \Sigma)}$, since it is the latter principle that is our proper subject. 

An attempt to derive Factorizability$_{t_P, 1, 2}$ from Local Causality$_{C(\Sigma, \Sigma)}$ has been made by \citet{ciepielewski-superdeterminism}. They propose decomposing $\lambda_{C(\Sigma, \Sigma)}$ into four parts (notation mine): $\lambda_{C(\Sigma, \Sigma), 1, 2}$ (which pertains to the measured system), $\lambda_{C(\Sigma, \Sigma), a}$ and $\lambda_{C(\Sigma, \Sigma),b}$ (which influence or determine the settings $a$ and $b$, respectively),  and $\lambda_{C(\Sigma, \Sigma),E}$ (i.e., everything else). 
Then, Factorizability$_{C (\Sigma, \Sigma)}$, 
\begin{equation}
P(A, B | a, b, \lambda_{C(\Sigma, \Sigma)}) = P(A | a,  \lambda_{C(\Sigma, \Sigma)})  P(B | b, \lambda_{C(\Sigma, \Sigma)}),
\end{equation}
becomes
\begin{equation}\label{eq:factorizability-Ciepielewski-1}
\begin{split}
& P(A, B | a, b, \lambda_{C(\Sigma, \Sigma), E}, \lambda_{C(\Sigma, \Sigma),a}, \lambda_{C(\Sigma, \Sigma),b}, \lambda_{C(\Sigma, \Sigma),1, 2}) = \\
& P(A | a, \lambda_{C(\Sigma, \Sigma), E}, \lambda_{C(\Sigma, \Sigma),a}, \lambda_{C(\Sigma, \Sigma),b}, \lambda_{C(\Sigma, \Sigma),1, 2})   P(B | b, \lambda_{C(\Sigma, \Sigma), E}, \lambda_{C(\Sigma, \Sigma),a}, \lambda_{C(\Sigma, \Sigma),b}, \lambda_{C(\Sigma, \Sigma),1, 2}).
\end{split}
\end{equation}
Now, assuming that (i) $\lambda_{C(\Sigma, \Sigma),E}$ is irrelevant, whereas (ii) $\lambda_{C(\Sigma, \Sigma),a}$ and $\lambda_{C(\Sigma, \Sigma),b}$ influence the outcome only via $a$ and $b$, which were already taken into account, they conclude that only $\lambda_{C(\Sigma, \Sigma),1, 2}$ remains relevant, so from \eqref{eq:factorizability-Ciepielewski-1} they obtain
\begin{equation}\label{eq:factorizability-Ciepielewski-2}
P(A, B | a, b, \lambda_{C(\Sigma, \Sigma), 1, 2}) = P(A | a, \lambda_{C(\Sigma, \Sigma), 1, 2})  P(B | b, \lambda_{C(\Sigma, \Sigma), 1, 2}).
\end{equation}

However, this reasoning is contestable. \citet[p.~11]{ciepielewski-superdeterminism} justify (i) by saying that $\lambda_E$ ``is irrelevant by definition''. This is true only if the measurement outcomes depend only on the measurement settings and the state of the system at the time of preparation. However, they might also be influenced by other factors, such as the microstates of the measurement devices or some details of the environment. Moreover, the state of the system might change between the time of preparation and the time of measurement, being influenced by $\lambda_{C(\Sigma, \Sigma),E}$ (among other things). Also (ii) might be false because the same factor can influence both the measurement settings and something else that influences the measurement outcomes (not via the settings).

Importantly, what they ultimately want to derive is not \eqref{eq:factorizability-Ciepielewski-2} but Factorizability$_{t_P, 1, 2}$, which is
\begin{equation}\label{eq:factorizability-Ciepielewski-3}
P(A, B | a, b, \lambda_{t_P 1, 2}) = P(A | a, \lambda_{t_P, 1, 2})  P(B | b, \lambda_{t_P 1, 2}).
\end{equation}
This is because for the proof of Bell's theorem to work, we need the same $\lambda$ in both Factorizability and Settings Independence. However, \eqref{eq:factorizability-Ciepielewski-2} and \eqref{eq:factorizability-Ciepielewski-3} are not equivalent. As already observed, the state of the measured system might change between the time of preparation and the time of measurement. Therefore, $\lambda_{C(\Sigma, \Sigma), 1, 2}$ might not be the same as $\lambda_{t_P, 1, 2}$ (and if position is one of the hidden variables, they surely will be different). Denoting the state of the system just by $\lambda$, without any indices, obscures this issue.

The above-mentioned problem with (i) was recognised by \citet{ciepielewski-locality}, which motivated them to make a different attempt to derive Factorizability$_{t_P, 1, 2}$ from Local Causality$_{C(\Sigma, \Sigma)}$ (cf. also their criticism of some previous derivations by other authors, pp. 6--14). This time, their starting point is not Factorizability$_{C (\Sigma, \Sigma)}$ but the following mathematical fact:
\begin{equation}\label{eq:ciepielewski-start}
P(A, B | a, b, \lambda_{t_P, 1, 2}) = 
P(A|B, a,b, \lambda_{t_P, 1, 2})  P(B| a, b, \lambda_{t_P, 1, 2}).
\end{equation}
The main idea of this new derivation is as follows \citep[p.~15]{ciepielewski-locality}:
\begin{quote}
[...] assuming the principle of local 
causality, under what circumstances would it be illegitimate to remove $b$ or $B$ from the first
term on the right-hand side of [\eqref{eq:ciepielewski-start}]? For that to be the case, two conditions are required. First, there must be something, besides $a$ and $\lambda$, which is relevant for the prediction. Second, that something must be (at least partially) encoded in $b$ or $B$.
\end{quote}
Then they claim that the first assumption is plausible, whereas the second amounts to the violation of what they call ``Microstate Independence'':
\begin{equation}\label{eq:microstate-independence}
P(\lambda_A |b, B) = P(\lambda_A) \text{ and } P(\lambda_B|a, A) = P(\lambda_B),
\end{equation}
where $\lambda_A$ is everything that influences the outcome $A$ besides $a$ and $\lambda_{t_P, 1, 2}$ (and analogously for $\lambda_B$). But what exactly does it mean to say that $\lambda_A$ is ``everything that influences the outcome $A$ besides $a$ and $\lambda_{t_P, 1, 2}$''? How to express this idea in terms of probabilities? \citet{ciepielewski-locality} do not give the answer. Perhaps this can be captured by the following condition, which is similar to Seevinck and Uffink's sufficiency condition:
\begin{equation}\label{eq:sufficiency-Ciepielewski}
\begin{split}
& P(A | a, b, B, \lambda_{C(\Sigma, \Sigma)}, \lambda_A, \lambda_{t_P, 1, 2}) = P(A | a, \lambda_A, \lambda_{t_P, 1, 2}) \text{ and }  \\
& P(B| a, b, A, \lambda_{C(\Sigma, \Sigma)}, \lambda_B, \lambda_{t_P, 1, 2}) = P(B | b, \lambda_B, \lambda_{t_P, 1, 2}).
\end{split}
\end{equation}
Therefore, instead of assuming the existence of $\overline{\lambda}$ that satisfies both Factorizability$_{\overline{\lambda}}$ and Settings Independence$_{\overline{\lambda}}$ (as in section \ref{sec:incomplete-sufficient}), we need to assume the existence of some $\lambda_A$ and $\lambda_B$ that satisfy \eqref{eq:microstate-independence} and \eqref{eq:sufficiency-Ciepielewski}. But are these assumptions more plausible? 

Even more importantly, does Factorizability$_{t_P, 1, 2}$ follow from these assumptions? \citet[p.~17]{ciepielewski-locality} claim that it does:
\begin{quote}
[...] the principle of local
causality implies that anything outside of the past light cone of 1, which does not enhance
the prediction for $A$ given $a$ and $\lambda$, can be removed from the conditional in the probability
of $A$ (and similarly for region 2). Then, we notice that, because of microstate independence,
this is the case for $b$ and $B$ in the first term on the right and for $a$ in the second, so
such terms can be removed, leading to factorizability. 
\end{quote}
However, Local Causality$_{C(\Sigma,\Sigma)}$ allows for removing factors from the conditional probability of $A$ only if there is another appropriate factor that we condition upon, which needs to cover the entire thick slice of the past light cone of $R_A$. This is not satisfied here because Eq. \eqref{eq:ciepielewski-start} does not involve any factor that covers the entire thick slice of the past light cone of $R_A$. Notice also that $\lambda_A$ does not appear anywhere in \eqref{eq:ciepielewski-start}, so it is not clear how \eqref{eq:microstate-independence} and \eqref{eq:sufficiency-Ciepielewski} could justify removing any terms from \eqref{eq:ciepielewski-start}.

\section{A new proposal}\label{sec:new-proposal}

The proposal for establishing \refbc{bell} that I want to put forward here is in some sense a combination of the second, third and fourth strategies listed in the second paragraph of section \ref{sec:improve}. It is the closest to the third strategy: we will derive something close to Factorizability$_{t_P, 1, 2}$ from Local Causality$_{C(\Sigma, \Sigma)}$. However, instead of $\lambda_{t_P, 1, 2}$, we will use a bit more encompassing state (the fourth strategy), and we will also appeal to a coarse-graining in order to bypass the problem of the possible changes of $\lambda_{t_P, 1, 2}$ in time (the second strategy).

Let us begin with the following observation. In principle, some factors different from the state of the two particles at $t_P$, the effects of which are encoded in their subsequent states, could explain the correlation between the outcomes of measurements made on them (cf. the fourth paragraph in section \ref{sec:derive-fact-from-lc}). But what else could these be? One can argue that all other factors, except for the arrangement of the measurement devices and the state-preparing device, could be different in every run of Bell's Experiment, so they cannot be responsible for the statistical regularities in the outcomes.
\citet[pp.~39--40]{butterfield-lewis-bell} formulates this idea in the following way:
\begin{quote}
Consider an experiment that gets repeated many
times. Its repetitions will not share all their properties; any two are liable
to differ in myriad ways (such as temperature and humidity). But it may
be that no property divides the class of repetitions into those with the
property and those without it so as to give different statistics in the two
subclasses. In such a case, call the experiment's statics ``homogeneous''. [...]
The more the repetitions vary among themselves while the statistics
nevertheless remain homogeneous, the more we expect that the properties common to all the repetitions are exactly all the properties on
which the chances of the phenomena at issue depend. 
\end{quote}

Therefore, when looking for the causes of the correlation between measurement outcomes, it is reasonable to restrict ourselves to the two-particle system at $t_P$, or at least to a close neighbourhood of it (in case it is the state-preparing device or something always present in its immediate vicinity rather than the two-particle system itself that causes the correlation).
We will call this state $\lambda_{t_P, 1, 2, +}$; it includes $\lambda_{t_P,1,2}$ and presumably something more, but it is located entirely within the intersection of the past light cones of $R_A$ and $R_B$ at time $t_P$.%
\footnote{One of the reviewers contested the novelty of this idea. I agree that many other authors might have had something similar in mind implicitly. However, I want to stress that this choice of the localisation of the states under consideration does not exhaust the content of my proposal. It is only the first step towards finding a set of premises that are physically reasonable and that, when added to Local Causality, enable one to derive an appropriate version of Factorizability without rendering the corresponding version of Settings Independence implausible. When I say ``derive'', I mean literally \emph{derive}---in a rigorous, logical sense---and not merely argue in a hand-waving manner. I am not aware of any such derivation in the literature (\citealt{hofer-szabo-SU} includes a rigorous derivation but with a different formulation of Local Causality; see Appendix \ref{app:hofer-szabo} for a comparison), and I believe that it is important to have one. First, Bell's Conclusion, if true, is one of the most significant results in the foundations of QM, so we should have at our disposal a fully clean argument for it. Even if most researchers in the field already believe that Bell's Conclusion follows from Bell's theorem, we should know \emph{exactly} how it follows (and if there is more than one such derivation, based on different sets of supplementary assumptions, then we should know at least one of them). Second, once we have the mentioned set of assumptions, we obtain an exhaustive catalogue of logically legitimate ways of denying Bell's Conclusion (namely, by rejecting one of these assumptions as either false or having a false presupposition), so that we can nip in the bud many pointless speculations concerning alleged overlooked ways of avoiding Bell's Conclusion (perhaps there are some unrecognised options, but they would still need to fall into one of the categories on our list). }

Let us turn to our argument. Assume for reductio that Local Causality$_{C (\Sigma, \Sigma)}$ holds (i.e., that some HVT satisfying Local Causality$_{C (\Sigma, \Sigma)}$ is true). From Local Causality$_{C (\Sigma, \Sigma)}$ we can derive Factorizability$_{C (\Sigma, \Sigma)}$ (see Appendix \ref{app:Fact-from-LC}), 
\begin{equation}\label{eq:factorizability-CSS-repeat}
P(A, B | a, b, \lambda_{C(\Sigma, \Sigma)}) = 
P(A | a,  \lambda_{C(\Sigma, \Sigma)}) P(B | b, \lambda_{C(\Sigma, \Sigma)}).
\end{equation}
Let $\Lambda_{1, 2, +}$ be the set of possible states of our enlarged system (including two particles and ``something more'') at a single time. Assume that there exists%
\footnote{It should be stressed that the assumption here is that there \emph{exists at least one} such coarse-graining, \emph{not} that \emph{any} coarse-graining satisfies (A1)--(A2). For this reason, we should not be worried about Simpson-like effects, which in our case would amount to possible violations of (A1)--(A2) by other coarse-grainings.}
a coarse-graining of $\Lambda_{1, 2, +}$, denoted by $\overline{\Lambda}_{1, 2, +}$, such that in Bell's Experiment,
\begin{enumerate}
\item [(A1)] the aspects of the state of our enlarged system captured by this coarse-graining do not change between the preparation and the measurement, that is,
\begin{equation}\label{eq:A1}
\forall_{t_0 \in [t_P, t_M]} \ \overline{\lambda}_{t_0, 1, 2 , +} = \overline{\lambda}_{t_P, 1, 2 , +},
\end{equation}
where $t_M$ is the time of measurement (cf. footnote \ref{fn:tM}); and 
\item [(A2)] those aspects of the state of our enlarged system are also the only aspects of $\lambda_{C(\Sigma, \Sigma)}$ that are relevant for probabilities of measurement outcomes conditioned on measurement setting and $\lambda_{C(\Sigma, \Sigma)}$; that is, for some $t''$ such that $t < t'' < t'$,
\begin{equation}\label{eq:plus-AB-CSS}
P(A, B | a, b, \lambda_{C(\Sigma, \Sigma)}) = 
P(A, B | a, b, \overline{\lambda}_{t'', 1, 2 , +}) , 
\end{equation}
\begin{equation}\label{eq:plus-A-CSS}
P(A | a,  \lambda_{C(\Sigma, \Sigma)}) = 
P(A| a, \overline{\lambda}_{t'', 1, 2 , +}) 
\end{equation}
and
\begin{equation}\label{eq:plus-B-CSS}
P(B |  b, \lambda_{C(\Sigma, \Sigma)}) = 
P( B |  b, \overline{\lambda}_{t'', 1, 2 , +}) 
\end{equation}
where $t$ and $t'$ are is as in Fig. \ref{fig:LC}.
\end{enumerate}
Since $t_P < t < t'' < t' < t_M$, from \eqref{eq:plus-AB-CSS}--\eqref{eq:plus-B-CSS} together with \eqref{eq:A1} it follows that
\begin{equation}\label{eq:plus-AB-CSS'}
P(A, B | a, b, \lambda_{C(\Sigma, \Sigma)}) = 
P(A, B | a, b, \overline{\lambda}_{t_P, 1, 2 , +}), 
\end{equation}
\begin{equation}\label{eq:plus-A-CSS'}
P(A | a,  \lambda_{C(\Sigma, \Sigma)}) = 
P(A | a, \overline{\lambda}_{t_P, 1, 2 , +})
\end{equation}
and
\begin{equation}\label{eq:plus-B-CSS'}
P(B |  b, \lambda_{C(\Sigma, \Sigma)}) = 
P( B |  b, \overline{\lambda}_{t_P, 1, 2 , +}).
\end{equation}

Now, if we apply \eqref{eq:plus-AB-CSS'} to the LHS of \eqref{eq:factorizability-CSS-repeat}, and \eqref{eq:plus-A-CSS'} together with \eqref{eq:plus-B-CSS'} to the RHS of \eqref{eq:factorizability-CSS-repeat}, we obtain Factorizability$_{\overline{t_P, 1, 2, +}}$, that is,
\begin{equation}\label{eq:factorizability-tilde}
P(A, B | a, b, \overline{\lambda}_{t_P, 1, 2 , +}) = 
P(A | a, \overline{\lambda}_{t_P, 1, 2 , +})  P(B | b, \overline{\lambda}_{t_P, 1, 2 , +}).
\end{equation}
If we additionally assume Settings Independence$_{\overline{t_P, 1, 2, +}}$,
\begin{equation}\label{eq:SI-plus-tilde}
P(\overline{\lambda}_{t_P, 1, 2 , +} | a,b) = P(\overline{\lambda}_{t_P, 1, 2 , +}) ,
\end{equation}
then we will have the full package of assumptions that are needed to derive Bell's inequality with $\lambda = \overline{\lambda}_{t_P, 1, 2 , +}$ (i.e., Factorizability$_{\overline{t_P, 1, 2 ,+}}$ and Settings Independence$_{\overline{t_P, 1, 2, +}}$). 
Since Bell's inequality is violated, 
either Local Causality$_{C (\Sigma, \Sigma)}$ is false or there is no coarse-graining $\overline{\Lambda}_{1, 2, +}$ that satisfies (A1), (A2) and Settings Independence$_{\overline{t_P, 1, 2 ,+}}$.

Why should we believe that such a coarse-graining exists? Let us begin with observing that Settings Independence$_{\lambda_{t_P, 1, 2 , +}}$, that is,
\begin{equation}\label{eq:SI-plus}
P(\lambda_{t_P, 1, 2 , +} | a,b) = P(\lambda_{t_P, 1, 2 , +}) ,
\end{equation}
is already plausible (even without taking a coarse-graining of $\lambda_{t_P, 1, 2 , +}$). The justification for Settings Independence$_{t_P, 1, 2 ,+}$ is the same as the justification for Settings Independence$_{\lambda_{t_P, 1, 2 }}$. The entire state $\lambda_{t_P, 1, 2 , +}$ is confined to the intersection of the light cones of $R_A$ and $R_B$ at $t_P$. Therefore, the past light cones of $R_a$ and $R_b$ include plenty of factors that belong to neither the region covered by $\lambda_{t_P, 1, 2 , +}$ nor its causal future, so they are able to interfere with the influence of $\lambda_{t_P, 1, 2 , +}$ on the measurement settings. After coarse-graining, this version of Settings Independence becomes even more plausible: if there were any factors in $\lambda_{t_P, 1, 2 , +} $ correlated with the choices of measurement settings, they could be lost by taking an equivalence class of such states, but no new correlations can be introduced in this way.%
\footnote{This part of my reasoning can be underpinned formally. Assume that $X$ and $X'$ are such that (i) $X \cap X'= \emptyset$, (ii) $P(X | Y) = P(X)$ and (iii) $P(X' | Y) = P(X')$. Let their coarse-grained counterpart be $\overline{X} = X \cup X'$. 
We have $P(\overline{X} | Y) = P(X \cup X' | Y) = P(X | Y) + P(X' | Y) = P(X) + P(X') = P(\overline{X})$ (where the second and the last equality comes from (i), whereas the third equality comes from (ii) and (iii)). This can be generalised to any $\overline{X}$ that consists of countably many events.}

What about (A1) and (A2)? An important role in their justification is played by common-cause-like intuitions and continuous-action intuitions.%
\footnote{Importantly, however, we do not assume that $\overline{\lambda}_{t_P, 1, 2 , +}$ is a common cause in the sense of satisfying Factorizability$_{\overline{t_P, 1, 2, +}}$ because this would render Local Causality$_{C (\Sigma, \Sigma)}$ redundant in the derivation of Bell's inequality, which would again block our attempt to establish \refbc{bell}.}
First, whatever is responsible for the \textit{correlation} between measurement outcomes should lie in the common past of the two measurements, that is, in the intersection of the past light cones of $R_A$ and $R_B$. Otherwise it could influence (in a local way) at most one of the measurement outcomes but not both; however, the correlation involves both outcomes. A factor that lies outside of the intersection of the past light cones of $R_A$ and $R_B$ could spoil the correlation between the two particles, but it cannot create any correlation between them because it can influence at most one of them. Second, we have observed (see the fourth paragraph of section \ref{sec:derive-fact-from-lc}) that the state of the two particles can change between $t_P$ and $t_M$; \textit{a fortiori} this is true about the state of the two particles and ``something more''. However, this aspect of $\lambda_{t_P, 1, 2,+}$ that is responsible for the correlation should better last unchanged until the measurements on both wings are made, since otherwise its influence on measurement outcomes would require some kind of action at a temporal distance (cf. \citealt{adlam-temporal-locality}).%
\footnote{\citet[p.~2]{adlam-temporal-locality} defines the condition of Temporal Locality in the image of Factorizability, not Local Causality$_{C (\Sigma, \Sigma)}$. However, we can easily find a formulation of Temporal Locality that is closer to the latter:
\begin{quote}
Temporal Locality$_{C(\Sigma, \Sigma)}$, fine-grained version: For any bounded region $R$, 
for any $t < R$ and any $t' <t$, for any $\lambda_R$ and $\lambda_{C(\Sigma_{R, t}, \Sigma_{R, t'})}$, for any bounded region $R''<t'$ and any $\lambda_{R''}$,
\begin{equation}
P(\lambda_R |  \lambda_{C(\Sigma_{R, t}, \Sigma_{R, t'})}, \lambda_{R''})  =
P(\lambda_R | \lambda_{C(\Sigma_{R, t}, \Sigma_{R, t'})}).
\end{equation}
\end{quote}
Temporal Locality does not play any official role in our reasoning that leads to \refbc{bell}, but the underlying idea is used in our justification of (A1) and (A2). Moreover, it is interesting for us because its violation makes it possible to reconcile the satisfaction of Factorizability$_{\overline{t_P, 1, 2 ,+}}$ with the violation of Factorizability$_{C (\Sigma, \Sigma)}$: we can explain this by assuming that $\overline{\lambda}_{t_P, 1, 2 , +}$ acts directly on measurement outcomes, without the mediation of anything in the region $C (\Sigma, \Sigma)$---that is, in a temporally nonlocal way.
}
The idea that causation between temporally distant events needs to always be mediated by intermediate events is known in the literature about Bell's theorem as ``action-by-contact'' \citep{evans-price-wharton-new-slant} or ``continuous action'' (\citealt{wharton-argaman-locally-mediated} and \citealt{adlam-retrocausality}). Notably, this idea seems to be present in \citeauthor{Bell1990}'s (\citeyear[p.~239]{Bell1990}) initial formulation of Local Causality (``[t]he direct causes (and effects) of events are near by''), but it is not reflected in Local Causality$_{C(\Sigma, \Sigma)}$, which does not prohibit events in the remote past but lying in the past-light cone of $R$ to influence the probability of a given state in $R$.

I do not know whether the existence of a coarse-graining that satisfies (A1) and (A2) can be derived in a formal way from some more basic principles, so I will leave the justification of this assumption in the above form.

\section{Summary}\label{sec:conclusion}

In this paper, an important ambiguity in the understanding of hidden states in the literature about Bell's theorem has been analysed. Hidden states in the context of Bell's theorem are usually considered either as the states of the two-particle system at the moment of preparation or as the states of thick slices of the past light cones of measurements. In both cases, there are problems with establishing \refbc{bell}, which states that all HVTs that satisfy Local Causality are inconsistent with the predictions of QM for Bell's Experiment. In the first approach, the link between Factorizability$_{t_P, 1, 2}$ and Local Causality$_{C(\Sigma, \Sigma)}$ is missing. In the second approach, Settings Independence$_{C(\Sigma, \Sigma)}$ is violated by all locally deterministic theories (unless we reject either Physical Nature of Human Choices or the Universality of Laws, but this also undermines the proof that all locally deterministic theories satisfy Local Causality$_{C(\Sigma, \Sigma)}$). Since the condition of Local Causality$_{C(\Sigma, \Sigma)}$ was intended by Bell to be a generalisation of local determinism, locally deterministic theories form a very important (even a paradigmatic one) subclass of locally causal theories. Therefore, the failure to establish \refbc{bell} for this class of theories cannot be ignored.

Three possible ways of improving this situation have been identified. One can stay with $\lambda = \lambda_{t_P, 1, 2}$ and try to derive Factorizability$_{t_P, 1, 2}$ from Local Causality$_{C(\Sigma, \Sigma)}$. This, however, requires some additional assumptions, and it is not clear what they should be. One can also stay with $\lambda = \lambda_{C(\Sigma, \Sigma)}$ and consider coarse-grained versions of these states. However, to establish \refbc{bell}, we need to assume that there exists a coarse-graining $\overline{\Lambda}$ for which both Factorizability$_{\overline{\lambda}}$ and Settings Independence$_{\overline{\lambda}}$ hold, and it is not clear how to justify this assumption. Finally, one can consider some other choices of $\lambda$. My own proposed approach combines the ideas of all three strategies.

Let me list what I think are the major results of this paper.
First of all, my considerations show that it \textit{does} matter where exactly the $\lambda$ that occurs in Bell's theorem is located. 
Second, I have pointed out that the relationship between deterministic theories and $0$-$1$ probabilities is not straightforward and I have proposed the Bridge Principle, which allows us to relate them (Definition \ref{def:bridge}). 
Third, I have provided a precise definition of locally deterministic theories (Definition \ref{def:locally-det-theory}) and shown that they are deterministic (Theorem \ref{thm:local-det-implies-det}), that they satisfy Local Causality$_{C(\Sigma, \Sigma)}$ (Theorem \ref{thm:local-det-LocalCausality}) and that they violate Settings Independence$_{C(\Sigma, \Sigma)}$ (Theorem \ref{thm:local-det-violationSettInd}); the latter two results are based on the aforementioned Bridge Principle. 
Fourth, this Bridge Principle has also allowed us to distinguish clearly between deterministic and outcome deterministic theories (see the third paragraph of section \ref{sec:state-of-system}). 
Finally, it seems that there is at least one plausible way of establishing \refbc{bell}; the one that I have presented relies on some additional assumptions that allow one to derive the relevant variant of Factorizability from Local Causality$_{C(\Sigma, \Sigma)}$ (see section \ref{sec:new-proposal}).

Assuming that the above considerations are correct, let us reflect on the scope of the conclusions that we can derive from Bell's theorem. 
If ``nonlocal'' means ``either locally indeterminate or locally determinate but violating Local Causality$_{C(\Sigma, \Sigma)}$'', then we can strengthen what we called ``\refbc{bell}'' as follows:
\begin{quote}
\textbf{Strengthened Bell's Conclusion:} All HVTs consistent with the predictions of QM for Bell's Experiment are nonlocal.
\end{quote}%
If a given HVT is locally indeterminate, this follows from the above definition of nonlocality; and if a given HVT is locally determinate, this follows from Bell's Conclusion.
Since Nature realises the predictions of QM for Bell's Experiment, 
if we assume that one of HVTs needs to be a true description of Nature,
we can conclude that Nature itself is nonlocal in the above sense. However, I want to emphasise that this concept of nonlocality is not just the violation of Local Causality$_{C(\Sigma, \Sigma)}$ but a disjunction whose first disjunct is an even deeper form of nonlocality---namely, the failure of local determinateness.%
\footnote{Can a theory (either the standard QM or an HVT) be nonlocal in both senses? I think that, strictly speaking, the answer is no. The condition of Local Causality, as formulated by late Bell and explicated in Definitions \ref{def:LC-fg} and \ref{def:LC-cg}, presupposes the framework of local beables, in which it makes sense to ascribe the values of variables to any region of spacetime---something directly contradicting local determinateness. Of course, one may consider generalisations of this framework that also allow locally indeterminate entities, but Local Causality formulated in such a generalised framework would---in my view---entirely lose its bite. Surely, if a locally indeterminate entity is present in two spacelike-related regions, then it can exert causal influence in both of them, thereby leading to causal correlations between spacelike-related events (e.g., measurement outcomes). However, what is strange here is not the causal action of that entity (it acts where it is present---no surprise!) but its local indeterminateness. \label{fn:nonlocal-two-senses}
}

\bibliographystyle{apalike} 

\appendix

\section{The derivation of the Bell-CHSH inequality from Factorizability and Settings Independence}\label{app:derivation-inequality}

Below I present a proof of the Bell-CHSH inequality that makes explicit the use of both Factorizability and Settings Independence.%
\footnote{This proof follows quite closely the presentation by \citet[pp.~237--240]{norsen-foundations-2017}; the main difference is that here it is made explicit at which point we need to use Settings Independence. 
}
Define the expectation value of the product of the outcomes of measurements in directions $a$ and $b$ by 
\begin{equation}
E(a,b ) =  \int \sum_{A,B} A B P (A,B | a,b,\lambda) P (\lambda | a, b) d \lambda .
\end{equation}

We will show that if Factorizability and Settings Independence are satisfied, then
a particular combination of such expectation values, $|E (a,b) + E (a,b')|  + |E (a',b) - E (a',b')|$, is less than or equal to $2$. Consider the first term:
\begin{equation}\label{eq:Bell-proof-1}
\begin{split}
\left| E (a,b) + E (a,b') \right| & = \\
\left|  \int \sum_{A,B} A B P (A,B | a,b,\lambda) P (\lambda | a, b) d \lambda 
+\int \sum_{A,B'} A B' P (A,B' | a,b',\lambda) P (\lambda | a, b') d \lambda \right| & = \\
\left|  \int \left( \sum_{A,B} A B P (A,B | a,b,\lambda) 
+ \sum_{A,B'} A B' P (A,B' | a,b',\lambda) \right) P (\lambda ) d \lambda \right| & = \\
\left|  \int \left( \sum_{A,B} A B P (A | a, \lambda) P(B | b, \lambda)
+ \sum_{A,B'} A B' P (A | a,\lambda) P(B' | b', \lambda) \right) P (\lambda ) d \lambda \right| & = \\
\left|  \int \sum_A A P (A | a, \lambda) \left( \sum_{B}  B  P(B | b, \lambda)
+ \sum_{B'}  B'  P(B' | b', \lambda) \right) P (\lambda ) d \lambda \right| & \leq \\
 \int \left| \sum_A A P (A | a, \lambda) \right| 
 \left| \sum_{B}  B  P(B | b, \lambda)
+ \sum_{B'}  B'  P(B' | b', \lambda)  \right| P (\lambda ) d \lambda & \leq \\
\int  \left| \sum_{B}  B  P(B | b, \lambda)
+ \sum_{B'}  B'  P(B' | b', \lambda)  \right| P (\lambda ) d \lambda
\end{split}
\end{equation}
The third line is obtained by using Settings Independence, the fourth by using Factorizability, the sixth by using the fact that $\left| \int f (x) dx \right| \leq \int |f(x) | dx$, $|y z| = |y| |z|$ and $P (\lambda) \geq 0$, and the seventh by using the fact that $| \sum_A A P (A | a, \lambda)| \leq 1$ (which holds because $A = \pm 1$ and $P (A | a, \lambda) \leq 1$). Analogously, we can show that the second term satisfies the following inequality:
\begin{equation}\label{eq:Bell-proof-2}
\left| E (a',b) - E (a',b') \right| \leq 
\int  \left| \sum_{B}  B  P(B | b, \lambda)
- \sum_{B'}  B'  P(B' | b', \lambda)  \right| P (\lambda ) d \lambda.
\end{equation}
By combining \eqref{eq:Bell-proof-1} and \eqref{eq:Bell-proof-2}, we get:
\begin{equation}
\begin{split}
\left| E (a,b) + E (a,b') \right|  + \left| E (a',b) - E (a',b') \right| &  \leq \\
\int \left( \left| \sum_{B}  B  P(B | b, \lambda)
+ \sum_{B'}  B'  P(B' | b', \lambda)  \right| 
+ \left| \sum_{B}  B  P(B | b, \lambda)
- \sum_{B'}  B'  P(B' | b', \lambda)  \right|  \right) P (\lambda ) d \lambda & \leq 2.
\end{split}
\end{equation}
The last inequality is obtained by using the mathematical fact that $|x+y| + |x-y| \leq 2$ for $|x| \leq 1$ and $|y| \leq 1$, and
$\int P (\lambda) d \lambda = 1$.

\section{Proofs}

This appendix collects proofs of the theorems from the main part of the paper.

\subsection{Locally deterministic theories are deterministic}\label{app:loc-det-implies-det}

\setcounter{theorem}{0}

\begin{theorem}
Any locally determinate theory that is locally deterministic is also deterministic.
\end{theorem}

\begin{proof}
Let $T$ be a locally determinate theory that is not deterministic.
This means that there exists time $t_0$ and state $\lambda_{t_0}$ such that two different solutions, $\lambda_{\text{spacetime}}  \neq \lambda'_{\text{spacetime}} $, are compatible with $\lambda_{t_0}$ and the laws of $T$. 
Since $\lambda_{\text{spacetime}}$ and $\lambda'_{\text{spacetime}} $ are different, there exists a spatiotemporal region $R$ on which they differ. That is, $\lambda_{\text{spacetime}}$ is compatible with some $\lambda_R$ and $\lambda'_{\text{spacetime}}$ is compatible with some $\lambda'_R$ such that $\lambda_R \neq \lambda'_{R}$. 
If region $R$ partially overlaps with $\Sigma_{t_0}$, then---since our spacetime is $\mathbb{R}^4$---there exists a bounded and connected region $R'$ such that $R' \subsetneq R$, $R'$ does not overlap with $\Sigma_{t_0}$, and the restrictions of $\lambda_{\text{spacetime}}$ and $\lambda'_{\text{spacetime}}$ to $R'$ are different (i.e., $\lambda_{R'} \neq \lambda'_{R'}$). Concerning the last condition, notice that $\lambda_{\text{spacetime}}$ and $\lambda'_{\text{spacetime}}$ by definition do not differ on $\Sigma_{t_0}$ but do differ on $R$, so they must differ on some part of $R$ that does not overlap with $\Sigma_{t_0}$, which is encompassed by some $R'$ satisfying the remaining conditions.
If $R$ does not overlap with $\Sigma_{t_0}$, let $R'$ be any bounded and connected subregion of $R$ on which $\lambda_{\text{spacetime}}$ and $\lambda'_{\text{spacetime}}$ differ. Since $R'$ does not overlap with $\Sigma_{t_0}$ and is bounded and connected, it must be wholly located either below or above $\Sigma_{t_0}$---that is, $t_0 < R'$ or $t_0 > R'$. 
Assume without loss of generality that $t_0 < R'$ (the other case is analogous).
Consider our initial conditions $\lambda_{t_0}$ restricted to the intersection of the past light cone of $R'$ with $\Sigma_{t_0}$---that is, $\lambda_{ \Sigma_{R', t_0}}$. Since there exist two different specifications of hidden variables in region $R'$ that are compatible with $\lambda_{ \Sigma_{R', t_0}}$ (namely, $\lambda_{R'}$ and $\lambda'_{R'}$), our theory $T$ is not locally deterministic. Therefore, $T$ not being deterministic entails that it is also not locally deterministic. By contraposition, if $T$ is locally deterministic, then it is also deterministic.
\end{proof}

\subsection{Derivation of Factorizability$_{C(\Sigma, \Sigma)}$ from Local Causality$_{C(\Sigma, \Sigma)}$}\label{app:Fact-from-LC}

We will derive some consequences for Bell's Experiment from the coarse-grained version of Local Causality$_{C(\Sigma, \Sigma)}$, which will enable us to prove Factorizability$_{C(\Sigma, \Sigma)}$.
Consider any times $t$ and $t'$ such that $t'<t$, $R_a$ and $R_b$ lie above $t$, and $t'$ is above the intersection of the past light cones of $R_A$ and $R_B$. Let $a$ and $b$ be possible experimental settings chosen in regions $R_a$ and $R_b$, and let $A$ and $B$ be their possible outcomes obtained in regions $R_A$ and $R_B$. By $a, b, A$ and $B$ we also denote the equivalence classes of $\lambda_{R_a}$, $\lambda_{R_b}$, $\lambda_{R_A}$ and $\lambda_{R_B}$, which correspond to the respective settings and outcomes. Additionally, $\lambda_{C(\Sigma, \Sigma)}$ denotes the pair consisting of $\lambda_{C(\Sigma_{R_A, t}, \Sigma_{R_A, t'})}$ and $\lambda_{C(\Sigma_{R_B, t}, \Sigma_{R_B, t'})}$, and $C(\Sigma, \Sigma)$ denotes the respective region. From Local Causality$_{C(\Sigma, \Sigma)}$ it follows that for any $\lambda_{C(\Sigma_{R_A, t}, \Sigma_{R_A, t'})}$ and any $\lambda_{C(\Sigma_{R_B, t}, \Sigma_{R_B, t'})}$,
\begin{equation}\label{eq:local-causality-exp-1}
P(A, a | B, b, \lambda_{C(\Sigma_{R_B, t}, \Sigma_{R_B, t'})}, \lambda_{C(\Sigma_{R_A, t}, \Sigma_{R_A, t'})})=
P(A, a |  \lambda_{C(\Sigma_{R_A, t}, \Sigma_{R_A, t'})}) .
\end{equation}
In more detail, to obtain Eq. \eqref{eq:local-causality-exp-1} we have used Eq. \eqref{eq:local-causality-coarse-grained} with the following substitutions: 
$R_A \cup R_a$ for $R$, $A$ and $a$ for $\overline{\lambda_{R, d_{R}}}$, 
$R_B \cup R_b \cup C(\Sigma_{R_B, t}, \Sigma_{R_B, t'})$ for $R'$, and finally,
$B$, $b$ and $\lambda_{C(\Sigma_{R_B, t}, \Sigma_{R_B, t'})}$ for $\overline{\lambda_{R', d_{R'}}}$.
Notice that under these choices, regions $R$ and $R'$ are bounded (as they must be) but not connected.

It also follows from Local Causality$_{C(\Sigma, \Sigma)}$ that for any $\lambda_{C(\Sigma_{R_A, t}, \Sigma_{R_A, t'})}$ and any $\lambda_{C(\Sigma_{R_B, t}, \Sigma_{R_B, t'})}$,
\begin{equation}\label{eq:local-causality-exp-4}
P( a | b, \lambda_{C(\Sigma_{R_A, t}, \Sigma_{R_A, t'})}, \lambda_{C(\Sigma_{R_B, t}, \Sigma_{R_B, t'})})=
P( a |  \lambda_{C(\Sigma_{R_A, t}, \Sigma_{R_A, t'})}) .
\end{equation}
Finally, it follows from Local Causality$_{C(\Sigma, \Sigma)}$ that for any $\lambda_{C(\Sigma_{R_A, t}, \Sigma_{R_A, t'})}$ and any $\lambda_{C(\Sigma_{R_B, t}, \Sigma_{R_B, t'})}$,
\begin{equation}\label{eq:local-causality-exp-5}
P(A, a | \lambda_{C(\Sigma_{R_A, t}, \Sigma_{R_A, t'})}) = 
P(A, a | \lambda_{C(\Sigma_{R_A, t}, \Sigma_{R_A, t'})}, \lambda_{C(\Sigma_{R_B, t}, \Sigma_{R_B, t'})})
\end{equation}
and 
\begin{equation}\label{eq:local-causality-exp-6}
P( a |  \lambda_{C(\Sigma_{R_A, t}, \Sigma_{R_A, t'})}) =
P( a |  \lambda_{C(\Sigma_{R_A, t}, \Sigma_{R_A, t'})}, \lambda_{C(\Sigma_{R_B, t}, \Sigma_{R_B, t'})}),
\end{equation}
where we harmlessly added the conditionalization on the second hidden state.

The above results can be used to prove Factorizability$_{C(\Sigma, \Sigma)}$:
\begin{equation}
\begin{split}
P(A, B | a, b, \lambda_{C(\Sigma, \Sigma)}) =
P(A, B | a, b, \lambda_{C(\Sigma_{R_A, t}, \Sigma_{R_A, t'})}, \lambda_{C(\Sigma_{R_B, t}, \Sigma_{R_B, t'})}) & = \\
\frac{P(A, B , a, b, \lambda_{C(\Sigma_{R_A, t}, \Sigma_{R_A, t'})}, \lambda_{C(\Sigma_{R_B, t}, \Sigma_{R_B, t'})})}{P( a, b, \lambda_{C(\Sigma_{R_A, t}, \Sigma_{R_A, t'})}, \lambda_{C(\Sigma_{R_B, t}, \Sigma_{R_B, t'})})} & = \\
\frac{P(A, a | B, b, \lambda_{C(\Sigma_{R_A, t}, \Sigma_{R_A, t'})}, \lambda_{C(\Sigma_{R_B, t}, \Sigma_{R_B, t'})}) P(B, b, \lambda_{C(\Sigma_{R_A, t}, \Sigma_{R_A, t'})}, \lambda_{C(\Sigma_{R_B, t}, \Sigma_{R_B, t'})})}{P( a | b, \lambda_{C(\Sigma_{R_A, t}, \Sigma_{R_A, t'})}, \lambda_{C(\Sigma_{R_B, t}, \Sigma_{R_B, t'})}) P(  b, \lambda_{C(\Sigma_{R_A, t}, \Sigma_{R_A, t'})}, \lambda_{C(\Sigma_{R_B, t}, \Sigma_{R_B, t'})})} & = \\
\frac{P(A, a | \lambda_{C(\Sigma_{R_A, t}, \Sigma_{R_A, t'})}) P(B, b, \lambda_{C(\Sigma_{R_A, t}, \Sigma_{R_A, t'})}, \lambda_{C(\Sigma_{R_B, t}, \Sigma_{R_B, t'})})}{P( a |  \lambda_{C(\Sigma_{R_A, t}, \Sigma_{R_A, t'})}) P(  b, \lambda_{C(\Sigma_{R_A, t}, \Sigma_{R_A, t'})}, \lambda_{C(\Sigma_{R_B, t}, \Sigma_{R_B, t'})})} & = \\
\frac{P(A, a | \lambda_{C(\Sigma_{R_A, t}, \Sigma_{R_A, t'})}, \lambda_{C(\Sigma_{R_B, t}, \Sigma_{R_B, t'})}) P(B, b, \lambda_{C(\Sigma_{R_A, t}, \Sigma_{R_A, t'})}, \lambda_{C(\Sigma_{R_B, t}, \Sigma_{R_B, t'})})}{P( a |  \lambda_{C(\Sigma_{R_A, t}, \Sigma_{R_A, t'})}, \lambda_{C(\Sigma_{R_B, t}, \Sigma_{R_B, t'})}) P(  b, \lambda_{C(\Sigma_{R_A, t}, \Sigma_{R_A, t'})}, \lambda_{C(\Sigma_{R_B, t}, \Sigma_{R_B, t'})})} & = \\
P(A | a,  \lambda_{C(\Sigma_{R_A, t}, \Sigma_{R_A, t'})}, \lambda_{C(\Sigma_{R_B, t}, \Sigma_{R_B, t'})})  \frac{P(B, b, \lambda_{C(\Sigma_{R_A, t}, \Sigma_{R_A, t'})}, \lambda_{C(\Sigma_{R_B, t}, \Sigma_{R_B, t'})})}{P(  b, \lambda_{C(\Sigma_{R_A, t}, \Sigma_{R_A, t'})}, \lambda_{C(\Sigma_{R_B, t}, \Sigma_{R_B, t'})})} & = \\
P(A | a,  \lambda_{C(\Sigma_{R_A, t}, \Sigma_{R_A, t'})}, \lambda_{C(\Sigma_{R_B, t}, \Sigma_{R_B, t'})})  P(B | b, \lambda_{C(\Sigma_{R_A, t}, \Sigma_{R_A, t'})}, \lambda_{C(\Sigma_{R_B, t}, \Sigma_{R_B, t'})}) & = \\
P(A | a, \lambda_{C(\Sigma, \Sigma)}) 
P(B | b, \lambda_{C(\Sigma, \Sigma)}).
\end{split}
\end{equation}
In the consecutive steps of this derivation we have used the following: 
definition of $\lambda_{C(\Sigma, \Sigma)}$;
definition of conditional probability $P(X | Y) = \frac{P(X, Y )}{P(Y )}$; 
formula $P(X, Y ) = P(X | Y) P(Y )$; \eqref{eq:local-causality-exp-1} and \eqref{eq:local-causality-exp-4}; \eqref{eq:local-causality-exp-5} and \eqref{eq:local-causality-exp-6};
theorem $\frac{P(X, Y |Z)}{P(Y | Z)} = P(X | Y, Z)$; definition of conditional probability; 
and finally,
definition of $\lambda_{C(\Sigma, \Sigma)}$.

\subsection{Locally deterministic HVTs satisfy Local Causality$_{C(\Sigma, \Sigma)}$ but violate Settings Independence$_{C(\Sigma, \Sigma)}$}\label{app:loc-det-LC-SI}

\setcounter{theorem}{1}

\begin{theorem}
If $T$ is a locally deterministic HVT, then it satisfies Local Causality$_{C(\Sigma, \Sigma)}$.
\end{theorem}

\begin{proof}
In a locally deterministic theory, $P(\lambda_R | \lambda_{C(\Sigma_{R, t}, \Sigma_{R,t'})})$ is equal to $0$ or $1$ because $\lambda_{\Sigma_{R, t}}$ uniquely determines the complete state in $R$ (we use our Bridge Principle Between Determinism and 0-1 Probability, Definition \ref{def:bridge}).
Therefore, adding any further conditionalization either does not change the value of probability or makes it ill-defined (the latter will happen if we add conditionalization on some $\lambda_{R''}$ that is inconsistent with $ \lambda_{C(\Sigma_{R, t}, \Sigma_{R,t'})}$).
In particular, adding conditionalization in the same way as in the formulation of the fine-grained version of Local Causality$_{C(\Sigma, \Sigma)}$ will not change the value of conditional probability of $\lambda_R$. 
Similarly, $P(\overline{\lambda_{R; d_R}} | \lambda_{C(\Sigma_{R, t}, \Sigma_{R,t'})})$ equals $0$ or $1$ for any coarse-grained state $\overline{\lambda_{R; d_R}}$.
This entails the coarse-grained version of Local Causality$_{C(\Sigma, \Sigma)}$.
\end{proof}

\begin{theorem}
If $T$ is a locally deterministic HVT, then it violates Settings Independence$_{C(\Sigma, \Sigma)}$.
\end{theorem}

\begin{proof}
Assume that $T$ is locally deterministic, and consider Bell's Experiment with the usual notation. Let $t''$ be such that $t' < t'' < t$.
Consider any states in regions $R_a$, $R_b$, $\Sigma_{R_A,t''}$ and $\Sigma_{R_B,t''}$, denoted by $\lambda_{R_a}$, $\lambda_{R_b}$, $\lambda_{\Sigma_{R_A,t''}} $ and $\lambda_{\Sigma_{R_B,t''}} $.
Since $T$ is locally deterministic, $\lambda_{\Sigma_{R_A,t''}} $ is compatible with exactly one state in $R_a$, and analogously for $\lambda_{\Sigma_{R_B,t''}} $ and $R_b$. Therefore, any pair of states 
$\{ \lambda_{\Sigma_{R_A,t''}} , \lambda_{\Sigma_{R_B,t''}} \}$ is compatible with exactly one pair of states $\{ \lambda_{R_a}, \lambda_{R_b} \}$. \textit{A fortiori}, any pair of states $\{ \lambda_{C(\Sigma_{R_A,t}, \Sigma_{R_A, t'})}, \lambda_{C(\Sigma_{R_B,t}, \Sigma_{R_B, t'})} \}$ is compatible with exactly one pair of states $\{ \lambda_{R_a}, \lambda_{R_b} \}$.

Assume without loss of generality that $\lambda_{C(\Sigma_{R_A,t}, \Sigma_{R_A, t'})}$ is consistent with $a$ but not with $a'$, and $\lambda_{C(\Sigma_{R_B,t}, \Sigma_{R_B, t'})}$ is consistent with $b$ but not with $b'$. Then, from our Bridge Principle Between Determinism and 0-1 Probability (Definition \ref{def:bridge}) it follows that $P ( \lambda_{C(\Sigma, \Sigma)} | a,b) \neq 0$ and $P (\lambda_{C(\Sigma, \Sigma)} | a', b') = 0$. Therefore, $P ( \lambda_{C(\Sigma, \Sigma)} | a,b)$ and $P (\lambda_{C(\Sigma, \Sigma)} | a', b')$ cannot both be equal to $P (\lambda_{C(\Sigma, \Sigma)})$, which is a violation of Settings Independence$_{C(\Sigma, \Sigma)}$.
\end{proof}

It is worth pointing out that Theorem \ref{thm:local-det-violationSettInd} cannot be strengthened to ``any deterministic HVT violates Settings Independence$_{C(\Sigma, \Sigma)}$'', where ``locally deterministic'' has been replaced with ``deterministic''. In the case of theories that are deterministic but not locally deterministic, it is the very first step of the proof that will not work. 
This is because in such a theory, for any $t$ below $R_a \cup R_b$,
\begin{equation}
P(\lambda_{R_a} | \lambda_{t} ) = 0/1 \text{ and } P(\lambda_{R_b} | \lambda_{t} ) = 0/1
\end{equation}
but not necessarily
\begin{equation}\label{eq:not-strengthened}
P(\lambda_{R_a} | \lambda_{\Sigma_{R_A,t}} ) = 0/1 \text{ and } P(\lambda_{R_b} | \lambda_{\Sigma_{R_B,t}} ) = 0/1.
\end{equation}
In other words, in a theory that is deterministic but not locally deterministic, all physical events in $R_a \cup R_b$ are uniquely determined by the full initial conditions at $t$, but not necessarily by the part of these initial conditions that is contained within $\Sigma_{R_A, t} \cup \Sigma_{R_B, t}$. However, this does not mean that the choices of measurement settings are more ``free'' in theories that are deterministic but not locally deterministic. This only reveals the fact that a certain kind of locality is built into the formulation of Settings Independence$_{C(\Sigma, \Sigma)}$: it states that the choices of measurement settings are independent of the conditions in the past \textit{restricted to the past light cone of measurement}, and not that the choices of measurement settings are independent of the conditions in the past, \textit{full stop}.

\section{Approach by Gábor Hofer-Szabó}\label{app:hofer-szabo}

The approach developed by \citet{hofer-szabo-SU} uses a different formulation of Local Causality than the one presented in section \ref{subsec:state-of-region-Bell}, although it is arguably equally close to the informal idea of local causality as expressed by \citet{Bell1990}. The advantage of Hofer-Szabó's approach is that it allows one to derive Factorizability with $\lambda$ covering a small region from Local Causality with $\lambda$ covering a large region region together with one additional assumption. 
Below I reproduce Hofer-Szabó's approach in my notation and simplify it a bit, so that it is easier to compare with other proposals discussed in sections \ref{sec:state-of-system}--\ref{sec:new-proposal}.

\begin{figure}
\centering
\includegraphics[width=\textwidth]{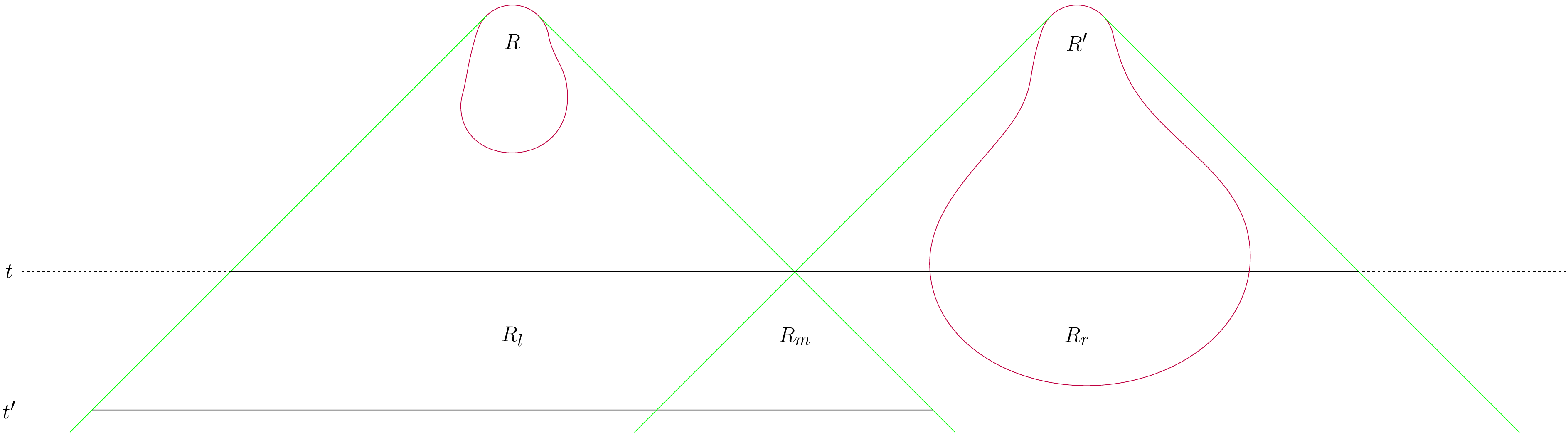}
\caption{Regions used in the formulation of Local Causality$_\text{H-S}$ and the derivation of Bell's theorem by \citet{hofer-szabo-SU}. Regions $R$ and $R'$ are within the purple lines.}\label{fig:Gabor}
\end{figure}

Let $t$ be such that it crosses the intersection of the past light cones of $R$ and $R'$, and let $t'< t$ (see Fig. \ref{fig:Gabor}, compare with Fig. \ref{fig:Bell-exp}). Region $C(\Sigma, \Sigma)$ does now include a part of the intersection of the light cones of  $R$ and $R'$.
Consider the division of region $C(\Sigma, \Sigma)$ into three disjoint parts: $R_m$ (in the intersection of the light cones of regions $R$ and $R'$), $R_l$ (in the light cone of $R$ but excluding its part that intersects the light cone of $R'$) and $R_r$ (in the light cone of $R'$ but excluding its part that intersects the light cone of $R$). It follows that $C(\Sigma, \Sigma) = R_m \cup R_l \cup R_r$ and $R_m \cap R_l = R_m \cap R_r = R_l \cap R_r = \emptyset$.
It is the state of the middle region, namely $R_m$, that will appear in the formulation of Bell's theorem.

The formulation of Local Causality to be used here is as follows:
\begin{definition}
Local Causality$_\text{H-S}$, fine-grained version: For any bounded region $R$ and any bounded region $R'$ that is space-like related to it, 
for any $t$ that crosses the intersection of the past light cones of $R$ and $R'$ at exactly one point, for any $t'<t$, and for any $\lambda_R$, $\lambda_{R'}$ and $\lambda_{C(\Sigma_{R, t}, \Sigma_{R, t'})}$, 
\begin{equation}\label{eq:local-causality-HS}
P(\lambda_R |  \lambda_{C(\Sigma_{R, t}, \Sigma_{R, t'})}, \lambda_{R'})  =
P(\lambda_R |  \lambda_{C(\Sigma_{R, t}, \Sigma_{R, t'})}).
\end{equation}
The coarse-grained version is analogous.
\end{definition}

The only difference between Local Causality$_\text{H-S}$ and Local Causality$_{C(\Sigma, \Sigma)}$ is that the former uses a thick slice that includes a part of the intersection of light cones of $R$ and $R'$, whereas the latter does not, in agreement with Bell's requirement that the thick slice used in the formulation of Local Causality ``completely shields off from [$R$] the overlap of the backward light cones of [$R$] and [$R'$]''.
However, the motivation that Bell gives for the proposed location of the thick slice is consistent with Hofer-Szabó's approach. Bell wanted the thick slice to be located above the intersection of the light cones of $R$ and $R'$ in order to avoid the situation in which there are some events above the thick slice but within that intersection such that they influence the outcomes of the experiment in a local way but are not uniquely determined by the state of the thick slice; the traces of such events might then be encoded in $\lambda_{R'}$, thereby making it relevant for the prediction of $\lambda_R$. However, if the upper part of the intersection is fully covered by the thick slice, as in Fig. \ref{fig:Gabor}, then such a situation cannot arise.

Notice that Local Causality$_\text{H-S}$ is consistent with part of region $R'$ lying below $t$ or even below $t'$, as long as $R'$ does not overlap with the causal past of $R$. For example, in Fig.~\ref{fig:Gabor}, $R'$ overlaps with $R_r$; it could even encompass the entire region $R_r$, which will be used below.

From Local Causality$_\text{H-S}$ it follows that
\begin{equation}\label{eq:HS-fromLC-1}
P(A, a | B, b , \lambda_l, \lambda_m, \lambda_r) = 
P(A, a |  \lambda_l, \lambda_m ) ,
\end{equation}

\begin{equation}
P( B, b | \lambda_l, \lambda_m, \lambda_r) = P( B, b |\lambda_m, \lambda_r) ,
\end{equation}

\begin{equation}
P( a | b, \lambda_l, \lambda_m, \lambda_r) = P( a |  \lambda_l, \lambda_m ) 
\end{equation}
and 
\begin{equation}\label{eq:HS-fromLC-4}
P(  b | \lambda_l, \lambda_m, \lambda_r) = P(  b | \lambda_m, \lambda_r).
\end{equation}

The additional assumption invoked by Hofer-Szabó is that the states of the three parts into which we have divided $C(\Sigma, \Sigma)$ are statistically independent:
\begin{equation}\label{eq:HS-Ind}
P(\lambda_l, \lambda_m, \lambda_r) = P(\lambda_l) P(  \lambda_m) P( \lambda_r).
\end{equation}

From Local Causality$_\text{H-S}$ and Eq. \eqref{eq:HS-Ind} once can derive Factorizability$_{\lambda_m}$ in the following steps:
\begin{equation}
\begin{split}
P(A, B | a, b, \lambda_m)
= \frac{P(A, B,  a, b,\lambda_m)}{P( a ,b, \lambda_m)} & = \\
\frac{\sum_{\lambda_l, \lambda_r} P(A, B,  a, b, \lambda_m, \lambda_l, \lambda_r) }{\sum_{\lambda_l, \lambda_r} P( a, b, \lambda_m, \lambda_l, \lambda_r)} & = \\
 \frac{\sum_{\lambda_l, \lambda_r} P(A, B,  a, b | \lambda_m, \lambda_l, \lambda_r) P(\lambda_m, \lambda_l, \lambda_r)}{\sum_{\lambda_l, \lambda_r} P( a, b | \lambda_m, \lambda_l, \lambda_r) P(\lambda_m, \lambda_l, \lambda_r)} & = \\
\frac{\sum_{\lambda_l, \lambda_r} P(A, a | B, b,  \lambda_m, \lambda_l, \lambda_r) 
P( B, b | \lambda_m, \lambda_l, \lambda_r) 
P(\lambda_m, \lambda_l, \lambda_r)}{\sum_{\lambda_l, \lambda_r} P( a | b, \lambda_m, \lambda_l, \lambda_r) 
P(  b | \lambda_m, \lambda_l, \lambda_r) 
P(\lambda_m, \lambda_l, \lambda_r)} 
& \overset{\text{\eqref{eq:HS-Ind}}}{=}  \\
\frac{\sum_{\lambda_l, \lambda_r} P(A, a | B, b, \lambda_m, \lambda_l, \lambda_r) 
P( B, b | \lambda_m, \lambda_l, \lambda_r) 
P(\lambda_m ) P( \lambda_l) P(\lambda_r)}{\sum_{\lambda_l, \lambda_r} P( a | b, \lambda_m, \lambda_l, \lambda_r) 
P(  b | \lambda_m, \lambda_l, \lambda_r) 
P(\lambda_m ) P( \lambda_l) P(\lambda_r)} 
& \overset{\text{\eqref{eq:HS-fromLC-1}--\eqref{eq:HS-fromLC-4}}}{=} \\
\frac{\sum_{\lambda_l, \lambda_r} P(A, a |  \lambda_m, \lambda_l ) 
P( B, b | \lambda_m,  \lambda_r) 
P(\lambda_m) P( \lambda_l) P(\lambda_r)}{\sum_{\lambda_l, \lambda_r} P( a |  \lambda_m, \lambda_l) 
P(  b | \lambda_m, \lambda_r) 
P(\lambda_m ) P( \lambda_l) P(\lambda_r)}
\frac{P(\lambda_m)}{P(\lambda_m )} & = \\
\frac{\sum_{\lambda_l} P(A, a |  \lambda_m, \lambda_l ) P(\lambda_m ) P( \lambda_l)}{\sum_{\lambda_l} P( a |  \lambda_m, \lambda_l) P(\lambda_m ) P( \lambda_l) }
\frac{\sum_{\lambda_r} P( B, b | \lambda_m, \lambda_r) P(\lambda_m ) P(\lambda_r)}{\sum_{\lambda_r} P(  b | \lambda_m, \lambda_r) P(\lambda_m ) P(\lambda_r)}
& =  \\
\frac{ P(A, a |  \lambda_m )P(\lambda_m ) }{ P( a |  \lambda_m) P(\lambda_m )  }
\frac{ P( B, b | \lambda_m ) P(\lambda_m ) }{ P(  b | \lambda_m  ) P(\lambda_m ) } & = \\
\frac{ P(A, a, \lambda_m  )}{ P( a, \lambda_m ) }
\frac{ P( B, b, \lambda_m)  }{ P(  b , \lambda_m ) } & = \\
P(A | a, \lambda_m )  P( B | b, \lambda_m )
\end{split}
\end{equation}

Since Settings Independence$_{\lambda_m}$ is plausible (for the same reason as Settings Independence$_{t_P, 1,2}$ and Settings Independence$_{\lambda_{t_P, 1, 2 , +}}$), we can run the proof of Bell's theorem with $\lambda = \lambda_m$.

\end{document}